\documentclass[structabstract]{aa}  
\usepackage{graphicx}
\usepackage{longtable,lscape}
\usepackage{float}
\usepackage{morefloats}
\restylefloat{figure}
\usepackage{txfonts}
\usepackage{natbib}
\bibpunct{(}{)}{;}{a}{}{,}
\begin{document}
   \title{A high-resolution spectroscopic survey of late-type stars: chromospheric 
   activity, rotation, kinematics and age\thanks{Based on observations made with:
   the 2.2m telescope of the German-Spanish Astronomical Centre, Calar Alto 
   (Almer\'{\i}a, Spain), operated jointly by the Max-Planck-Institute for Astronomy,
   Heidelberg, and the Spanish National Commission for Astronomy; 
   the Nordic Optical Telescope (NOT),
   operated on the island of La Palma jointly by Denmark, Finland,
   Iceland, Norway and Sweden, in the Spanish Observatorio del 
   Roque de Los Muchachos of the Instituto de Astrof\'{\i}sica de Canarias; 
   the Isaac Newton Telescope (INT)
   operated on the island of La Palma by the Isaac Newton Group in
   the Spanish Observatorio del Roque de Los Muchachos of the
   Instituto de Astrof\'{\i}sica de Canarias;
   with the Italian Telescopio Nazionale Galileo (TNG) 
   operated on the island of La Palma by the Centro Galileo Galilei of the 
   INAF (Istituto Nazionale di Astrofisica) at the Spanish Observatorio del 
   Roque de Los Muchachos of the Instituto de Astrof\'{\i}sica de Canarias;
   and with the Hobby-Eberly Telescope (HET)
   operated by McDonald Observatory on behalf 
   of The University of Texas at Austin, the
   Pennsylvania State University, Stanford University, 
   Ludwig-Maximilians-Universit\"at M\"unchen, and
   Georg-August-Universit\"at G\"ottingen.  This research has made use 
   of the SIMBAD database and VizieR catalogue access tool, operated at 
   CDS, Strasbourg, France.}}
   

   \subtitle{}

   \author{J. L\'opez-Santiago\inst{1}
                 \and 
                 D. Montes\inst{1}
                 \and
                 M. C. G\'alvez-Ortiz\inst{2}
                 \and
                 I. Crespo-Chac\'on\inst{1}
                 \and
                 R. M. Mart\'{\i}nez-Arn\'aiz\inst{1}
                 \and
                 M. J. Fern\'andez-Figueroa\inst{1}
                 \and
                 E. de Castro\inst{1}
                 \and
                 M. Cornide\inst{1}
          }
          
  \offprints{J. L\'opez-Santiago}

   \institute{Departamento de Astrof\'{\i}sica y Ciencias de la 
              Atm\'osfera, Universidad Complutense de Madrid,
              E-28040 Madrid, Spain\\
              \email{jls@astrax.fis.ucm.es}
         \and
              Centre For Astrophysics Research, 
              University of Hertfordshire, College Lane, Hatfield, 
              Hertfordshire, AL10 9AB, UK\\
             }

   \date{Received ...; accepted ...}

   \titlerunning{High-resolution spectroscopic survey of late-type stars}

 
  \abstract
   {}
   {We present a compilation of spectroscopic data from a survey of 
   144 chromospherically active young stars in the solar neighborhood
   which may be used to investigate different aspects of the formation
   and evolution of the solar neighborhood in terms of kinematics and 
   stellar formation history. The data have already been used by us in several 
   studies. With this paper, we make all these 
   data accessible to the scientific community for future studies on different 
   topics.}
   {We performed spectroscopic observations with \textit{echelle} spectrographs to 
   cover the entirety of the optical spectral range simultaneously. Standard data
   reduction was performed with the IRAF \textsc{echelle} package. We applied the
   spectral subtraction technique to reveal chromospheric emission in the stars 
   of the sample. The equivalent width of chromospheric emission lines was 
   measured in the subtracted spectra and then converted to fluxes using
   equivalent width--flux relationships. Radial and rotational velocities were 
   determined by the cross-correlation technique. Kinematics, equivalent widths of 
   the lithium line $\lambda6707.8$~\AA\ and spectral types were also determined.}
   {A catalog of spectroscopic data is compiled: radial and rotational velocities, space motion, 
   equivalent widths of optical chromospheric activity indicators from 
   \ion{Ca}{ii} H \& K to the calcium infrared triplet and the lithium line in 
   $\lambda$6708 \AA. Fluxes in the chromospheric emission lines
   and $R'_\mathrm{HK}$ are also determined for each observation of 
   star in the sample. {We used these data to investigate the emission levels
   of our stars. The study of the H$\alpha$ emission line revealed the presence of 
   two different populations of chromospheric emitters in the sample, clearly 
   separated in the $\log F_\mathrm{H\alpha}/F_\mathrm{bol}$ -- ($V-J$) diagram. 
   The dichotomy may be associated with the age of the stars.}}
   {}


   \maketitle
%

\section{Introduction}

The velocity (or phase) space in the 
solar neighborhood is rather complicated. In particular, 
the $UV$-plane shows different structures that are associated with the Galactic potential 
characteristics. At large scales, the $UV$-plane is dominated by long branches
\citep{sku99} related to dynamical perturbations altering the kinematics of the solar 
neighborhood \citep{fam05}, including the resonance of the rotating bar \citep{fam07,ant09}.
The fine structure of the velocity distribution of disk stars is more likely 
related to the existence of the classic Eggen moving groups \citep[see][for a review of the
young moving groups problem]{mon01a}. In fact, the substructures found inside
the long branches appear to have, on average, different ages \citep{asi99,ant08}, which 
agree with the idea of the moving groups being formed by coeval stars. 
%
Such young substructures are mixed in phase space with old stars and it is 
difficult to discern between young and old stars only by their 
kinematics. In some cases, the vertical component of the Galactic velocity ($W$) 
can be used to reject membership of the star in one of the young moving groups. 
The vertical velocity dispersion in the solar vicinity
is only dependent on the scale height \citep{tot92}. Old stars ($2-10$~Gyr) present 
higher scale heights than young stars ($< 1$~Gyr) and, hence, large values of $W$ are 
more typical of old stars. However, the division between high and low velocity is 
subtle.

{From the classical point of view of the Galactic velocity ellipsoid, 
early-type stars show lower dispersion in their Galactic velocities 
($U$, $V$, and $W$) than late-type stars and are usually restricted to the spiral arms
or star-forming regions. This result had already been interpreted in the past 
as a consequence of the increase of dispersion with increasing age \citep[see][]{mih81}. 
A recent study of the kinematics of M dwarfs in the solar vicinity by \citet{boc05} shows 
that the most chromospherically active (i.e. youngest) stars really show lower velocity 
dispersion than non-active (i.e older) stars. Because of this and their ubiquity, late-type 
stars are excellent tracers of the Galactic potential \citep{boc07}.
Distinguishing between young and old stars is then crucial to an understanding of the kinematics 
of the Galaxy and of stellar formation history in our neighborhood \citep{lop07}.
}

{An effort has been made in the past to quantify the proportion of young and old stars 
in samples of candidates of the young moving groups and associations using different age 
indicators. In particular, the level of magnetic activity is a powerful tool for this purpose 
\citep[see][]{lop09}. The activity level of late-type stars is inversely correlated with age 
due to the decrease of stellar rotation with increasing age \citep[e.g.][]{sku72,noy84,
rut87,ran96,piz03}. The rotation-age-activity relationship persists into the 
fully convective regime \citep[e.g.][]{rei07, wes08}. M dwarfs indeed have finite active 
lifetimes \citep{wes09iau}. Therefore, the level of magnetic activity 
is a good indicator of age for late-F to M dwarfs.
}
 
For four years, between 1999 and 2002, our group carried out a spectroscopic survey
of chromospherically active late-type stars in the solar neighborhood, selected from a list 
of possible members of classical young stellar kinematic groups \citep{mon01a}. 
The aim was to use different age indicators (chromospheric activity, lithium, rotational 
velocity) to constrain some properties of the moving groups and to study in detail the 
existence of age subgroups in large samples of stars selected, mainly, by their kinematics.
Since then, the information derived from this project has been extensively used by us in 
different works. For instance, \citet{mon01b} carried out a multi-wavelength study of 
a sample of active stars which allowed us to constrain the age of 14 young late-type stars.
{The results put constraints on the age of some young moving groups.}
Also, \citet{lop03} studied the relation between variations observed in the photosphere 
and chromosphere of \object{PW~And} using data from this spectroscopic 
survey. 

More recently, part of the data from this survey was used to investigate 
nearby young moving groups \citep{lop06}. A consequence
of this study was the confirmation of the existence of two age subgroups in the 
previously discovered AB Dor moving group \citep{zuc04}. 
{Using age indicators (mainly chromospheric activity and lithium abundance),
we showed that the Local Association is indeed a mixture of subgroups of stars with 
different ages.} In \citet{lop09}, 
we used the information on the chromospheric activity of the stars provided by the 
spectroscopic survey, together with X-ray data from the ROSAT All Sky Survey (RASS), 
to quantify the contamination by old main-sequence stars of the sample of possible 
members of the Local Association in \citet{mon01a}.

At present, two projects are being progressed by our group based on the stars in this survey: 
studying the connection between
various chromospheric activity indicators and the star formation process in the 
chromosphere \citep[see some preliminary results in][]{lop05, cre05}; and determining 
abundances of different elements in each star of the sample.

{In this paper, we compile data derived by us for the stars observed in our survey. 
Our aim is to make the spectroscopic data accessible to the scientific community for 
future studies. In this new era of Virtual Observatory and large 
photometric databases and catalogs, the compilation of spectroscopic data is important
for the purpose of constraining the properties of different astronomical objects, in particular, of stars 
\citep[e.g.][]{sci95,tak07,mic07,lop07,klu08,lop09}. The utility of large spectroscopic 
compilations has been demonstrated in the past by various groups. For instance, 
the Geneva-Copenhagen group derived metallicities, ages and kinematics from 
spectroscopic observations of a very large sample of FGK stars of the solar neighborhood
\citep{nor04}. These data were then used to investigate relations between metallicity, 
kinematics and age in the context of the evolution of the Galactic disk 
\citep{nor04,hol07,hol09}. Also, \citet{fuh04,fuh08} determined spectroscopic parameters
(temperature, gravity, metallicity, mass) of FGK stars in the solar vicinity with the aim of
constructing an unbiassed sample of stars in the Galactic thin and thick disk.
Similarly, \citet{all04} constructed a catalog of metallicities of late-type stars less than 
25 pc from the sun. 
}

{Several spectroscopic surveys were performed mainly to study the evolution of magnetic 
activity with age in late-type stars, or simply to investigate chromospheric activity in general. 
Thus, the Palomar/MSU group compiled a large sample of nearby M stars and determined 
both kinematics \citep{rei95} and chromospheric activity \citep{haw96}. The data were used, 
as well, to study the star formation history and luminosity function of the solar 
neighborhood \citep{giz02,rei02}. Another study on chromospheric emission is that of 
\citet{rau06} who measured \ion{Ca}{ii} H \& K in a large sample of K7--M stars.
Some groups are carrying out studies on how variations in line profiles produced
by chromospheric activity affect the detection of planets. These studies are producing 
new catalogs of natural targets of planet searches with chromospheric emission 
measurements \citep[e.g.][and references therein]{mar09}.
}

{Other surveys have specifically focused on determining spectroscopic 
parameters of active stars. For instance, \citet{str00} presents measurements 
of equivalent widths of \ion{Ca}{ii} H \& K, H$\alpha$, and \ion{Li}{i}, in addition to the 
kinematics of a large sample of active and inactive FGK stars. Also, \citet{tor06}
and \citet{gui09} determined kinematics and age indicators in large samples of 
late-type stars to select members of young moving groups and associations. 
\citet{whi07}
determined the same parameters for stars in the \textit{Spitzer} Legacy Science 
Program ``The Formation and Evolution of Planetary Systems'', aimed at studying
the formation and evolution of protoplanetary disks. And \citet{shk09} performed 
a spectroscopic survey of young M dwarfs within 25 pc. 
As \citet{val05} demonstrated in their spectroscopic analysis of effective temperature, 
surface gravity, metallicity, projected rotational velocity and abundance of 1040 
nearby FGK stars, the use of automated tools provides uniform results and makes the 
analysis of large samples practical.
}

{In this work, we determine 
equivalent widths and fluxes of most of the chromospheric activity indicators 
from \ion{Ca}{ii} H \& K to the \ion{Ca}{ii} infrared triplet (including the Balmer 
series and the \ion{Na}{i} doublet and \ion{Mg}{i} triplet) in a sample of stars with 
spectral types F to M. In this sense, our study implies an extension in terms of 
both spectral type range and wavelength coverage. We use the 
subtraction technique to subtract the photospheric contribution from the 
observed spectra, avoiding the use of calibrations to
determine chromospheric emission. This approach represents an advance 
on previous studies. 
Together with the kinematics, rotation, and equivalent widths of \ion{Li}{i} 
determined in this work, the sample constitutes an excellent laboratory for 
understanding the formation and evolution of the solar neighborhood during 
the past billion years. 
}

The structure of the paper is as follows: in Sect.~2, we give details of sample 
selection, observations, and data reduction methodology. In Sect.~3, we describe 
how we determined each parameter from the spectra. A brief summary of the results
is given in Sect.~4. The appendix contains tables with all the data and some 
figures with the spectra of stars in terms of selected chromospheric and photospheric features.


\section{Sample selection, observations and data reduction}
\label{sec2}

Our sample contains a total of 144 late-type stars. 
We selected 105 single stars from \citet{mon01a}. Due to their membership 
{in any of the moving groups studied in \citet{mon01a}
{(see $\S$~\ref{UVW} for a more detailed 
discussion on the stellar kinematic groups)},} we were quite certain that they are young 
and present chromospheric emission lines. 
{Although late-type K and M stars remain active for a long period
of their life \citep[e.g.][]{wes08}, adding the condition of being member of a young 
moving group places a greater restriction on a star's age. 
Nevertheless, since the \citet{mon01a} sample also contains some old  
stars \citep{lop09}, several stars of our
present work could be older than $\sim 1$ Gyr. 
}
The remaining 39 stars of our sample were 
selected because {they showed} a high level of magnetic activity, rotational rate, and/or lithium 
abundance. A complete list of the stars observed by us is given in Table~\ref{tab1}.
The sample is restricted in declination since the observations were taken in 
telescopes sited in the Northern Hemisphere. The minimum declination reached is
approximately $-20^\circ$. Another restriction is the brightness of the star. 
Since \textit{echelle} spectrographs have low efficiency, only stars with approximately 
$V \leq 12$~mag were selected. Our aim was to obtain spectra with high $S/N$ 
even for the faintest stars ($70-200$ in the region of H$\alpha$).
Spectral-type reference stars and radial velocity standards, used to 
determine the chromospheric excesses and heliocentric velocities
of our stars, were also observed in each campaign. To obtain robust 
results, we ensured good spectral-type coverage in each observing
run. A complete list of these stars is given in Table~\ref{tabA1}.

\begin{table*}[t]
\caption[]{Late-type stars studied in this work (HD number or other name). 
\label{tab1}}
\scriptsize
\begin{tabular}{lllllllll}
\hline \hline
\noalign{\smallskip}
            166  &            1405  &       1326  &            1835  &            2410  &          QT And  &            4568  &            4614  &          4614 B  \\
      BD+17 232  &           12230  &           13382  &           16525  &           17190  &           17382  &           17925  &           17922  &           18632  \\
          18803  &           20678  &           21845  &           23232  &           24916  &           25457  &           25680  &           25998  &           25665  \\
          29697  &           30652  &           33564  &           36869  &           37394  &          233153  &           41593  &   TYC 1355-75-1  &      BD+20 1790  \\
      HIP 39721  &        GJ 9251B  &       HIP 39896  &           72905  &           73171  &           77191  &           77407  &           82558  &           82443  \\
      HIP 47176  &       HIP 49544  &       HIP 50156  &          GJ 388  &       HIP 51317  &           85270  &           98736  &         GJ 426B  &          102392  \\
         105631  &          238087  &          238090  &          106496  &       HIP 60661  &          110010  &       HIP 62686  &       HIP 63023  &          112542  \\
         112733  &          115043  &       HIP 65016  &          238224  &          117860  &       HIP 67092  &         125161B  &          129333  &          133826  \\
         134319  &          135363  &          140913  &          142764  &          143809  &          145675  &          146696  &         147379A  &         147379B  \\
      HIP 79796  &          149661  &          149931  &          152863  &          152751  &         155674A  &         155674B  &          156984  &       HIP 84794  \\
      HIP 85665  &          160934  &          162283  &       HIP 87579  &       HIP 87768  &         GJ 698B  &          165341  &         GJ 702B  &          167605  \\
         234601  &        SAO 9067  &          168442  &       HIP 89874  &          171488  &          171746  &          173739  &          173740  &   2RE J1846+191  \\
        GJ 734B  &          184525  &          187458  &          187565  &          191011  &      HIP 101262  &          197039  &      HIP 102401  &          198550  \\
         200560  &          200740  &          201651  &      HIP 104383  & EUVE J2113+04.2  &      HIP 105885  &      HIP 106231  &          205435  &          206860  \\
         208472  &      HIP 108467  &      HIP 108752  & TYC1680-01993-1  &          209458  &      HIP 109388  &        V383 Lac  &         GJ 856B  &          213845  \\
     BD+17 4799  &      HIP 112460  &          216899  &          217813  &      HIP 114066  &          220140  &          221503  &      HIP 117779  &      HIP 118212  \\
\noalign{\smallskip}
\hline
\end{tabular}
\end{table*}

The observations were carried out during twelve observing runs between 1999 and 
2002. We used high resolution \textit{echelle} spectrographs (resolving power, 
$\lambda/\Delta\lambda$, ranging from 30,000 to 60,000 at 6500~\AA, $\Delta\lambda \sim 0.15$ \AA), 
with the exception of one observing run, where we used a long-slit spectrograph with 
spectral resolution $\Delta\lambda = 1.13$ \AA. Eighteen stars were observed 
with the latter configuration, {nine of them were observed only during this observing run. 
For these nine stars, we determined only radial and rotational velocities and measured 
equivalent widths of \ion{Li}{i} when possible. In general, errors in the measurements are 
only slightly larger for these stars than for those observed with higher resolution 
(see Table~\ref{tabA2}). However, the low resolution of the spectra in this observing
run prevented us from measuring chromospheric emission in these nine stars to compare
with our high resolution spectra \citep[see][for a discussion on this issue]{wal09}.}
Details of each observing run are given in Table~\ref{tab2}: date,
telescope, spectrograph, CCD chip, spectral range covered, number of orders
included in each \textit{echelle} spectrum, range of reciprocal dispersion and
spectral resolution (determined as the full width at half maximum, 
FWHM, of the arc comparison lines).
Some of the stars were observed more than once (in the same campaign or even 
in different ones). The total number of spectra collected for this survey is 518 for 
targets and more than 50 for standards.

\begin{table*}
\caption[]{Observing run details.
\label{tab2}}
\scriptsize
\begin{tabular}{lllllllll}
\noalign{\smallskip} \hline \hline \noalign{\smallskip}
Id. & \multicolumn{1}{c}{Date} & \multicolumn{1}{c}{Telescope} & 
Instrument & CCD chip & Spectral range & Orders & 
\multicolumn{1}{c}{Dispersion} & \multicolumn{1}{c}{FWHM}  \\
     &   &   &   &   & \multicolumn{1}{c}{(\AA)} & & \multicolumn{1}{c}{(\AA)} & \multicolumn{1}{c}{(\AA)} \\
\noalign{\smallskip} \hline \noalign{\smallskip}
1 &24-29 07/1999  & 2.2m$^{\rm a}$  & FOCES$^{\rm 1}$   & 2048x2048 15$\mu$m LORAL\#11 & 3910 - 9075  &  84 & 0.03 - 0.07 & 0.09 - 0.15 \\
2 &26-27 11/1999 & NOT$^{\rm b}$    & SOFIN$^{\rm 2}$       & 1152x770 EEV P88200       & 3525 - 10425 &  44 & 0.06 - 0.17 & 0.14 - 0.32 \\
3 &18-22 01/2000  & INT$^{\rm c}$    & MUSICOS$^{\rm 3}$ & 1024x1024 24$\mu$m TEK5        & 4430 - 10225 &  73 & 0.07 - 0.15 & 0.16 - 0.30 \\
4 &05-11 08/2000   & INT$^{\rm c}$    & MUSICOS$^{\rm 3}$ & 1024x1024 24$\mu$m TEK5      & 4430 - 10225 &  73 & 0.07 - 0.15 & 0.16 - 0.30 \\
5 &10-13 11/2000 & NOT$^{\rm b}$    & SOFIN$^{\rm 2}$       & 1152x770 EEV P88200               & 3525 - 10425 &  44 & 0.06 - 0.17 & 0.14 - 0.32 \\
6 &02-05 04/2001     & INT$^{\rm c}$    & IDS$^{\rm 4}$         & 2148x4200 13.5$\mu$m EEV10a      & 3554 - 7137  &   1 & 0.48        & 1.22        \\
7 &21-24 09/2001& 2.2m$^{\rm a}$  & FOCES$^{\rm 1}$       & 2048x2048 24$\mu$m Site\#1d   & 3510 - 10700 & 112 & 0.04 - 0.13 & 0.08 - 0.35 \\
8 &10-11 10/2001 & TNG$^{\rm d}$ & SARG$^{\rm 5}$  & 2(2048x4096) 13.5$\mu$m EEV 4280 & 4960 - 10110 &  62 & 0.02 - 0.04 & 0.08 - 0.17 \\
9 &19 12/2001 -  & HET$^{\rm e}$ & HRS$^{\rm 6}$ & 2(2048x4096) 15$\mu$m Marconi & 5040 - 8775 &  52 & 0.06 - 0.11 & 0.15 - 0.28 \\
   &28 02/2002    & & & & & & & \\
10& 22-25 04/2002 & 2.2m$^{\rm a}$  & FOCES$^{\rm 1}$   & 2048x2048 24$\mu$m Site\#1d  & 3510 - 10700 & 112 & 0.04 - 0.13 & 0.08 - 0.35 \\
11& 01-06 07/2002    & 2.2m$^{\rm a}$  & FOCES$^{\rm 1}$  & 2048x2048 24$\mu$m Site\#1d & 3510 - 10700 & 112 & 0.04 - 0.13 & 0.08 - 0.35 \\
12& 21-29 08/2002   & NOT$^{\rm b}$   & SOFIN$^{\rm 2}$  & 2048x2048 2K3EB PISKUNOV1  & 3525 - 10200 &  42 & 0.02 - 0.05 & 0.05 - 0.15 \\
\noalign{\smallskip} \hline \noalign{\smallskip}
\end{tabular}

{\scriptsize $^{\rm a}$ 2.2~m telescope at German Spanish
Astronomical Observatory (CAHA) (Almer\'{\i}a, Spain).\\
$^{\rm b}$ 2.56~m Nordic Optical Telescope (NOT) at Observatorio del Roque
de los Muchachos (La Palma, Spain).\\
$^{\rm c}$ 2.5~m Isaac Newton Telescope (INT) at Observatorio del Roque de
los Muchachos (La Palma, Spain).\\
$^{\rm d}$ 3.5~m Telescopio Nazionale Galileo (TNG) at Observatorio del
Roque de los Muchachos (La Palma, Spain).\\
$^{\rm e}$ 9.2~m Hobby-Eberly Telescope (HET) at McDonald Observatory (Texas, USA).\\
}
\vspace{-0.1cm}

{\scriptsize
$^{\rm 1}$ FOCES: Fiber Optics Cassegrain Echelle Spectrograph.\\
$^{\rm 2}$ SOFIN: Soviet Finnish High Resolution Echelle Spectrograph.\\
$^{\rm 3}$ MUSICOS: spectrograph developed as part of MUlti-SIte COntinuous Spectroscopy project.\\
$^{\rm 4}$ IDS: Intermediate Dispersion Spectrograph.\\
$^{\rm 5}$ SARG: Spettrografo di Alta Resoluzione Galileo.\\
$^{\rm 6}$ HRS: High Resolution Spectrograph.
 }
\end{table*}

For data reduction, we used the standard procedures in the
IRAF\footnote{IRAF is distributed by the National Optical Observatory,
which is operated by the Association of Universities for Research in
Astronomy, Inc., under contract with the National Science Foundation.}
package (bias subtraction, extraction of scattered light
produced by optical system, division by a normalized flat-field 
and wavelength calibration). After reduction, each spectrum
was normalized to its continuum, order by order, by fitting a polynomial 
function. 

\section{Results}
\label{sec3}

\subsection{Radial velocities}

Heliocentric radial velocities were determined using the cross-correlation
technique. In each observing run, the spectrum of each star was
cross-correlated order by order against spectra of radial velocity standards 
of similar spectral type (stars marked with an asterisk in Table~\ref{tabA1})
using the routine \texttt{fxcor} in IRAF. 
For each observed spectrum, radial velocities were derived for different spectral 
orders from the position of the peak of the cross-correlation function (CCF) by 
fitting a Gaussian to the function. Then, weighted means were calculated with 
the individual values obtained for each spectral order.
To avoid systematic errors produced by the effect of cool spots in the CCF, 
we fitted the Gaussian to the entire CCF profile, instead of fitting only the peak. 

Our results are listed in Table~\ref{tabA2}. We give radial velocities for each 
observation of the star ($V_\mathrm{r}$) and a mean velocity determined 
from the individual results for each observation ($\overline{V}_\mathrm{r}$).
Although our sample was selected from a list of single stars, some of 
them are actually single-line spectroscopic binaries. Known binaries in our
sample are: \object{HD 16525}, \object{HD 17190}, \object{HD 17382}, 
\object{HD 140913}, \object{HD 167605}, \object{HIP 89874}\footnote{HIP 89874
(FK Ser) is a binary with a separation of 1.33 arcsec \citep{her88,jen96}. In our 
spectra, we were not able to separate the two components.} and 
\object{HD 208472}. 
During our observations, other stars presented variations in their radial 
velocities that were hardly attributable to the presence of spots. 
If caused by spots, they 
should produce noticeable asymmetries in the absorption line 
profiles and this was not observed in their spectra. For other stars, we 
determined radial velocities which were quite different from those given in the literature.
Therefore, we classified all these stars as possible binaries: \object{BD+28 1779},
\object{HD 85270}, \object{GJ 466}, \object{HD 112542}, \object{HD 112733},
\object{HD 238224}, \object{HD 160934}, \object{BD-05 5480}, and 
\object{GJ 842.2}. \object{HD 160934} was confirmed as a binary system by
\citet{hor07} with preliminary orbital parameters determined by
\citet{gal06}.

In stars with a large coverage of the
stellar disk by spots, the asymmetries that they produce in the CCF are 
large enough to induce variations in the radial velocity measurements of 
up to several kilometers per second, 
even when a fit to the entire cross-correlation profile is performed 
\citep{dem92,str00,lop03}. We observed variations of this order for several 
stars in our sample for which we took spectra during different nights of
the same observing run: \object{HD 1405}, \object{BD+17 232}, 
\object{BD+20 1790}, \object{HD 72905}, \object{HD 82558}, \object{AD Leo}, 
\object{HD 135363}, \object{HD 171488}, \object{V383 Lac} and
\object{HD 220140}.

In Table~\ref{tabA2}, we also give the photometric periods available in the 
literature for some of the stars in the sample.

\subsection{Space motion}
\label{UVW}

Galactic space-velocity components ($U$, $V$, $W$) were 
determined as in \citet{mon01a}, who used a
modified version of the original procedure of \citet{joh87} to 
calculate Galactic velocities {and associated uncertainties.
As in \citet{mon01a}, we do not correct ($U$, $V$, $W$) 
for solar motion to make comparisons
with other works regarding moving groups easier.
} 
We used Hipparcos and Tycho-2 data \citep{esa97,hog00} and 
radial velocities determined by us. 
{For the ten stars with no available distance measurements 
in the literature, we determined a spectroscopic parallax using 
information on spectral type and luminosity class from our spectra.
The \citet{sch82} color--magnitude relations were used to determine 
$M_\mathrm{V}$ for these stars.
Note that for late-K and M stars, classic relations are not appropriate  
for determining some observed quantities. In particular, 
better spectroscopic parallax relations have been developed in the literature
using molecular bands \citep[e.g.][]{boc05}. Nevertheless, the spectral types of 
the stars in our sample for which we have determined a spectroscopic parallax
are in the range F5-K3. Note also that small variations in the parallax 
($< 10$~mas) of stars produce only small variations in the Galactic
velocities ($< 0.1$~km\,s$^{-1}$). 
} 

The resultant $U$, $V$ and $W$ velocity components are listed in 
Table~\ref{tabA2}. We used the mean radial velocity determined by us 
($\overline{V}_\mathrm{r}$ in the table) for this computation. For the 
possible binary systems, we did not correct for binarity since their 
orbits are yet unknown. Thus, we used the observed radial velocities 
--\,or the mean value when more than one observation was 
performed\,-- as a first approximation. 

As we mentioned in $\S$~\ref{sec2}, most of the stars in our sample 
(105) were selected from \citet{mon01a} who give a list of members and
possible members of the young moving groups. For them, we obtained 
Galactic velocity components very similar to those given in 
\citet{mon01a}. The remaining 39 stars of our sample had no previous 
measurement of $UVW$ velocities. As a first estimate, from their position 
in the {$UV$-- and $WV$-plane}, {50 stars could be classified as 
members or possible members of the Local Association (in its various subgroups), 
25 of the Hyades Supercluster, 20 of the Ursa Major Moving Group, 9
of the IC2391 Supercluster, 5 of the Castor Moving Group, and the other 17 
as young disk stars with no clear membership (see Table~\ref{tabA2}). 
The same velocity dispersion as in \citet{mon01a} was used for determining
the membership of the stars in any of the moving groups. We refer the reader
to that paper for a detailed explanation. 
Surprisingly, 18 stars are well outside
the classical boundaries of the young disk population in the $UV$-plane.
Most of them are the lowest active stars in our sample, but two of them are very
active stars: \object{HIP 79796} and \object{HD 216899} (see 
Fig.~\ref{figA1}-\ref{figA4}).
}

\subsection{Rotational velocities}

To determine the rotational velocities in our sample, we used a methodology
based on the cross-correlation technique, as we did for radial velocities
\citep[see][for a detailed discussion]{sod89}. The width of the peak of the CCF 
depends on the physical processes contributing to the line profile. 
The mathematical concept is very similar to that of the convolution of a 
theoretical spectrum with a rotation profile \citep[see][for details]{gra05}.
But, instead of comparing the stellar spectrum with a rotationally broadened 
one, the cross-correlation is performed between the spectrum of the 
\textit{program} star and a non-rotating one observed with the same instrument. 
Best results are obtained when both the comparison and the {program} stars 
have similar spectral types.

\begin{figure}[t!]
   \centering
   \includegraphics[width=8.3cm]{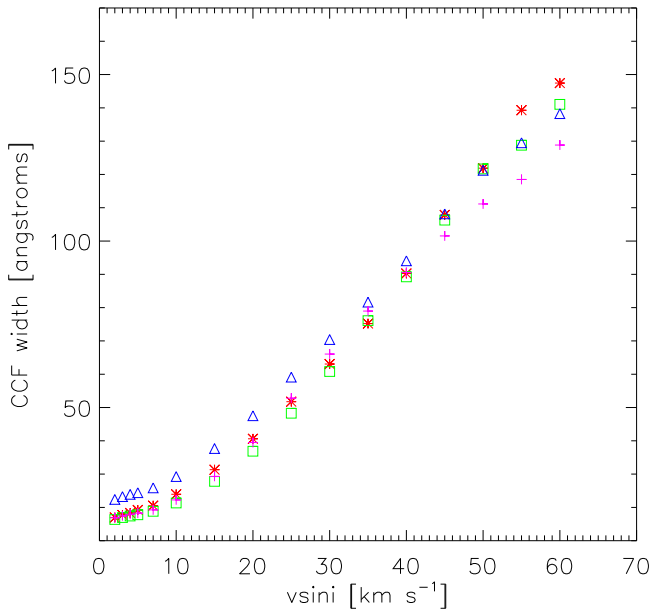} 
   \caption{CCF width--$v \sin i$ relation for standard stars of different 
   spectral type in the same observing run: HD~182488 (G8V, asterisks), 
   HD~185144 (K0 V, squares), HD~166620 (K2V, plusses), and 
   HD~201091 (K5V, triangles).}
   \label{f0}
\end{figure} 

{First}, for each reference (non-rotating) star observed with our sample, 
we calibrated the relation between the CCF width and $v \sin i$ {value}
by cross-correlating {the star with itself after rotationally broadening its spectrum 
at different velocities (with values ranging from 1 to 60 km\,s$^{-1}$). The result
is a relation between CCF width and rotational velocity (see Fig.~\ref{f0}).
The relation depends on the spectral type of the star (as shown in the figure). 
Then, we cross-correlated the program star with a standard with similar spectral 
type and used the relation for this standard to determine the rotational velocity
of the program star. Since the relation also depends on the instrumental 
configuration, the cross-correlation was performed separately in each observing run. 
A detailed explanation of the method can be found in \citet{lop03}.} 
%
For our study, the CCF peak width was determined by fitting a 
Gaussian function to it. This method ensures good results for 
$v \sin i \le 50$ km\,s$^{-1}$ \citep{sod89}. In our sample, there is only one 
star with $v \sin i > 50$ km\,s$^{-1}$ (\object{LO Peg}). For this star, we 
determined its rotational velocity by comparison with artificially broadened 
spectra of different stars with similar spectral types.  

In Table~\ref{tabA2}, we give each value of $v \sin i$ obtained for the stars in 
the sample and a mean value determined from the individual results. 
Uncertainties were determined using the parameter $R$ defined by 
\citet{ton79} as the ratio of CCF height to the $rms$ antisymmetric component. 
This parameter is computed by \texttt{fxcor} and provides a measurement of the 
signal-to-noise ratio of the CCF. \citet{ton79} showed that errors in the CCF width 
are proportional to  $(1 + R)^{-1}$, while \citet{har86} and \citet{rho01} 
found that the quantity $\pm v \sin i (1 + R)^{-1}$ provides a good estimate for 
the $90\%$ confidence level of the $v \sin i$ measurement. Thus, we 
adopted $\pm v \sin i (1 + R)^{-1}$ as a reasonable estimate of the 
uncertainties in our determinations.

\subsection{Spectral types and the lithium line}

During each observing run, a number of spectral-type standards were observed, 
covering the range of spectral types in our sample (from F to M). 
{To determine spectral types, we performed fits of our sample stars to 
spectral-type standards using a modified version of \textsc{starmod} \citep[][see
$\S$~\ref{sec_activity} for a detailed description of the procedure]{bar85}. 
The software first rotationally broadens the spectrum of the standard star 
until the best fit is obtained. Then, it subtracts the obtained synthetic spectrum 
from the sample star. In theory, if both stars have the same spectral type, 
the resultant (subtracted) spectrum should be null. In practice, the subtracted
spectrum shows some noise, due to small differences in metallicity and/or gravity
and also when the S/N of one of the spectra is low. Nevertheless, small differences in 
metallicity and gravity are lower than those produced by the difference of one
spectral subtype. 
The procedure of fitting is repeated with each one of the standard stars until
the best result is obtained. Errors are estimated in one spectral subtype. 
} 

\begin{figure}[t!]
   \centering
   \includegraphics[width=8.6cm]{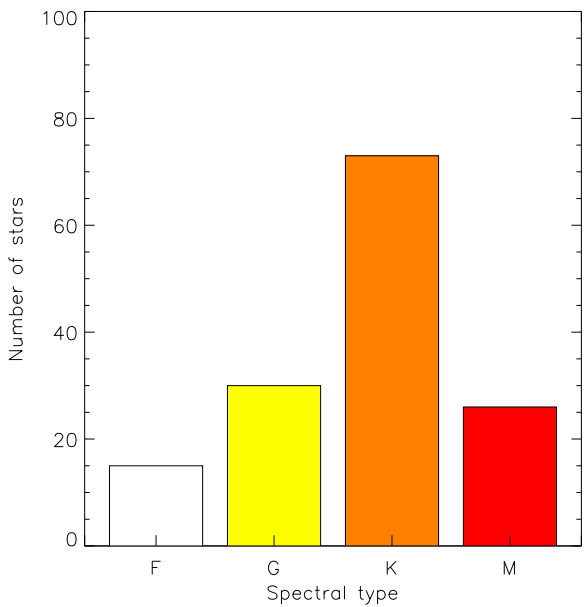} 
   \caption{Spectral type distribution of the stars in the sample.}
   \label{f1}
\end{figure}

Lines sensitive to spectral type were also used to determine spectral type. 
In particular, we used the lines \ion{Fe}{i} $\lambda$6430 \AA, \ion{Fe}{ii}
$\lambda$6432 and 6457 \AA, \ion{Ca}{i} $\lambda$6449 and 6456 \AA, 
\ion{Co}{i} $\lambda$6455 \AA, and \ion{V}{i} $\lambda$6452 \AA, as 
described in \citet{str90}. Other spectral lines used in this work for spectral 
classification are the \ion{Mg}{i} triplet $\lambda$5167, 5172, and 5183 \AA,
the \ion{Na}{i} doublet $\lambda$5590 and 5596 \AA, 
\ion{Ca}{i} $\lambda$6573 \AA\ and \ion{Fe}{i} $\lambda$6575 \AA.
{In contrast to the subtraction technique, this method is suitable 
only for slow rotators, since the lines involved in each relation of \citet{str90} 
are blended in stars with large rotational velocities. Note that the relations are 
calibrated only for FGK stars, but not for M ones. We used this method to 
test the results obtained with the subtraction technique for the stars in 
our sample showing small values of $v \sin i$. The largest differences 
are two spectral subtypes for F stars and one subtype for G and K stars.
These values are inside the uncertainties of the relations of \citet{str90}.
Our final results are given in Table~\ref{tabA2} (see also Fig.~\ref{f1}). 
}


We measured the equivalent width of the lithium line at 6707.8 \AA\ in the
observed spectra. The lithium abundance is an appropriate age indicator for 
approximately $\log age \leq 8.8$ Myr (the age of the Hyades cluster) since 
this element is easily destroyed by thermonuclear reactions in the stellar 
interior. Thus, detection of the lithium line is generally a sign of stellar 
youth in single stars (in binaries, high rotation velocities could preserve 
the lithium from depletion over longer time scales). 
{The lithium-age relation is mass-dependent. The 
usual way to determine age range is by comparison with stars of 
similar spectral type in clusters of well-known age 
\citep[e.g.][]{sod93b,neu97,mon01b}.}
In our spectra, the 
\ion{Li}{i} $\lambda$6707.8 \AA\ line is blended with the 
\ion{Fe}{i} $\lambda$6707.4 \AA\ {line}. To correct the measured equivalent
width ($EW$[\ion{Li}{i}+\ion{Fe}{i}]) for the \ion{Fe}{i} line, we used the
empirical relationship of \citet{sod93}. The results are given in 
Table~\ref{tabA2}. An error-weighted mean value of the individual 
$EW$[\ion{Li}{i}] measured on different nights and over different 
observing runs was also determined (see Table~\ref{tabA2}). 
Many stars in the sample still show the \ion{Li}{i} absorption line in 
their spectra (see Table~\ref{tabA2} and Fig.~\ref{figA3}).
Some of them are very young ($\log age \leq 7.5$ Myr). They belong 
to the young stellar associations \citep{zuc04,tor08}. In particular, 
\citet{lop06} established different age subgroups in the AB Dor 
moving group using results from this study.

\subsection{Chromospheric activity}
\label{sec_activity}

A special feature of \textit{echelle} spectrographs is that they cover a
large fraction of the optical spectrum. It allows simultaneous 
observations of all the activity indicators in this spectral range to be obtained.
%
In this work, we measured equivalent widths of the chromospheric optical lines: 
\ion{Ca}{ii} H \& K, H$\epsilon$, H$\delta$, H$\gamma$, H$\beta$, H$\alpha$, 
and the \ion{Ca}{ii} infrared triplet, for each observation of each star in our 
sample, when the spectrograph configuration permitted it. In very active stars
(active M dwarfs and flare stars), we also measured equivalent widths of the
lines: \ion{He}{i} $\lambda$5876 \AA\ and the \ion{Na}{i} doublet 
($\lambda\lambda$5890, 5896 \AA).

To remove the photospheric contribution, we used the spectral subtraction 
technique \citep[see details in][]{mon00}. 
{The advantage of this method is that no assumption about the 
continuum value for the equivalent width measurement is needed.}
Reference (non-active) stars with similar spectral types to our 
targets were used as templates (see Table~\ref{tabA1}). The subtraction was 
performed with \textsc{jstarmod}, a modified version of the Fortran code \textsc{starmod} 
developed at the Pennsylvania State University \citep{hue84,bar85}. Our modifications
permit the program to use \textit{echelle} spectra in the file format given by the 
majority of observatories. \textsc{jstarmod} first selects the region of the 
spectrum indicated by the user. Then, it rotationally broadens and shifts the
template spectrum to fit the target one. Finally, it subtracts the synthetic spectrum
from the observed one. 
{If both stars --\,active and non-active\,-- are identical in terms of 
photosphere, the resultant subtracted spectrum is the chromospheric emission 
of the active star: i.e. a flat spectrum with emission features at the positions of the
chromospheric lines. In practice, the subtracted spectrum shows some noise 
away from the chromospheric lines, due to small differences in metallicity and/or gravity. 
}

{A source of uncertainties is the possible basal emission of the non-active 
stars used as templates. Its consequence is a reduction in the measured 
equivalent widths of the \ion{Ca}{ii} H \& K chromospheric emissions of the target. 
However, for the subtraction we used reference stars situated 
close to the lower boundary of the surface flux in \ion{Ca}{ii} H \& K of 
\citet{rut84}. We estimate the largest uncertainty in logarithm fluxes as 0.1 dex, 
due to basal chromospheric emission from standard non-active stars.
%
}

The equivalent widths were determined by fitting a Gaussian function to 
the emission line profiles. For the \ion{Ca}{ii} H and H$\epsilon$ lines, 
which are blended in our spectra, we used a double Gaussian fit.
To obtain an estimation of the errors, we followed the
methodology explained in \citet{lop03}. Our results are given in Table~\ref{tabA3}.
Each measurement of the equivalent width and its uncertainty is listed in the table, 
together with the observing run and the Modified Julian Date (MJD) of the 
observation. Equivalent widths were later converted to absolute chromospheric 
fluxes at the stellar surface using the calibrations of \citet{hal96} (see results in
Table~\ref{tabA4}).  

\begin{figure*}[t!]
   \centering
   \includegraphics[width=6cm]{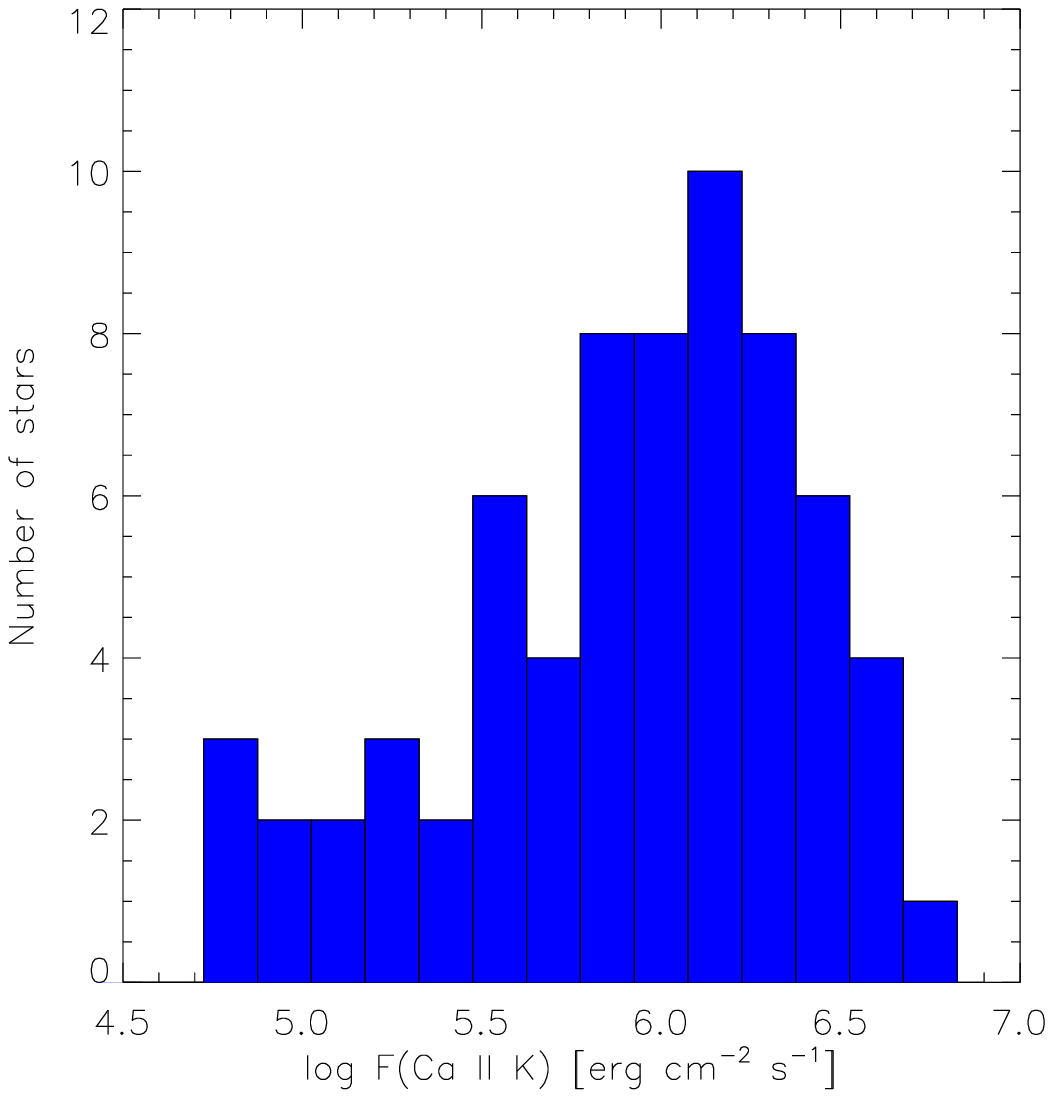}
   \includegraphics[width=6cm]{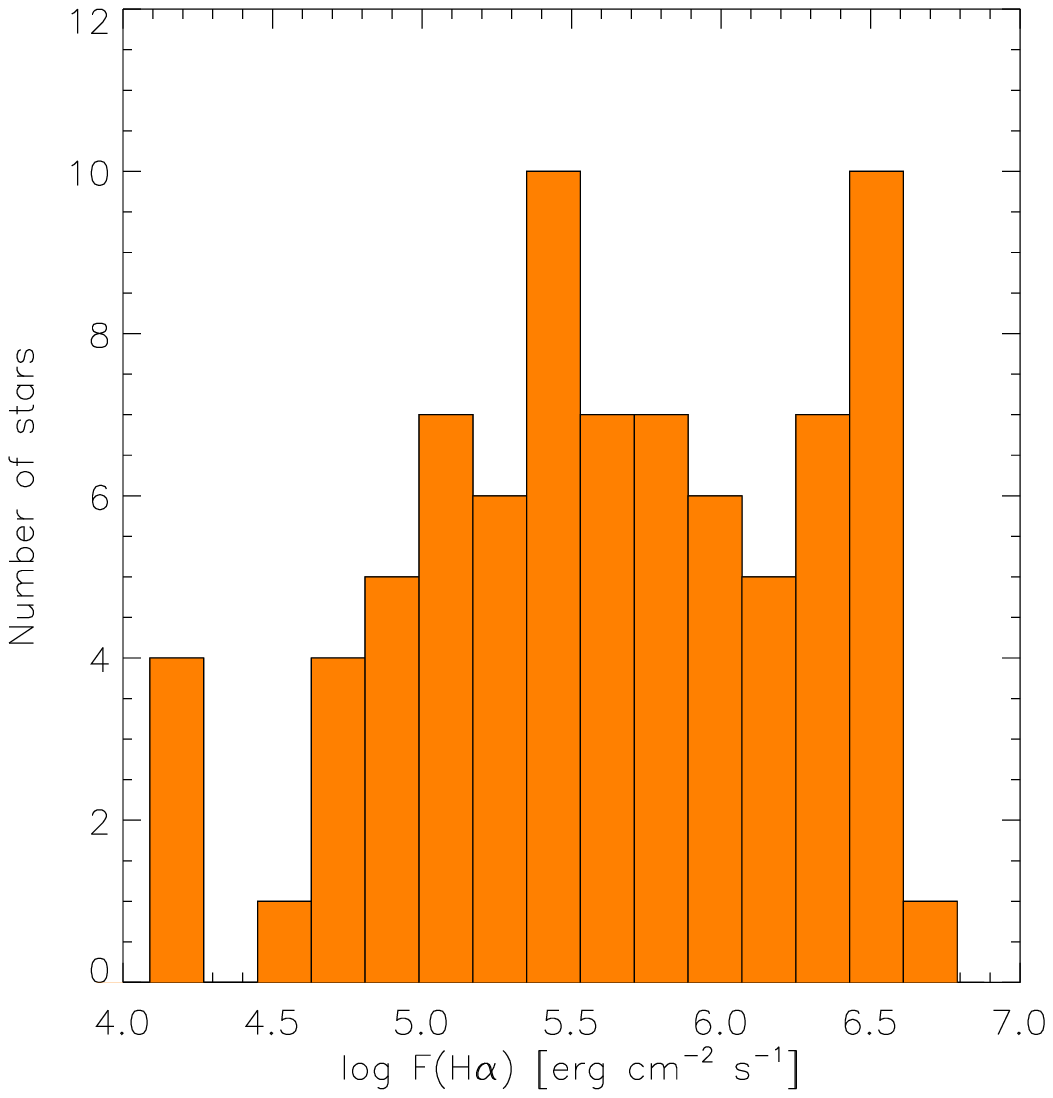}
   \includegraphics[width=6cm]{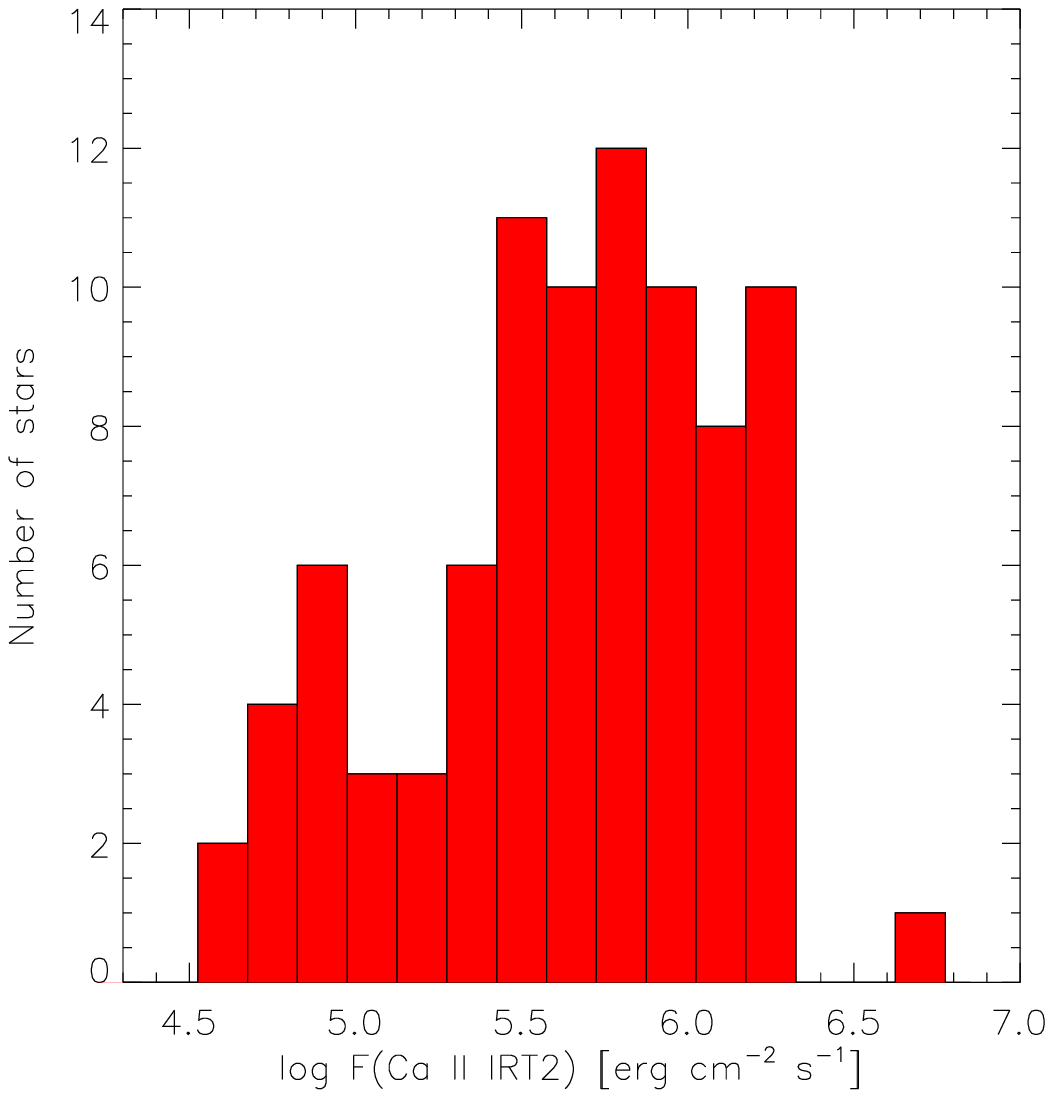}
   \caption{Histograms for results on some chromospheric activity
   indicators (absolute flux after subtraction of the photosphere).}
   \label{f2}
\end{figure*}

\begin{figure*}[t!]
   \centering
   \includegraphics[width=8.3cm]{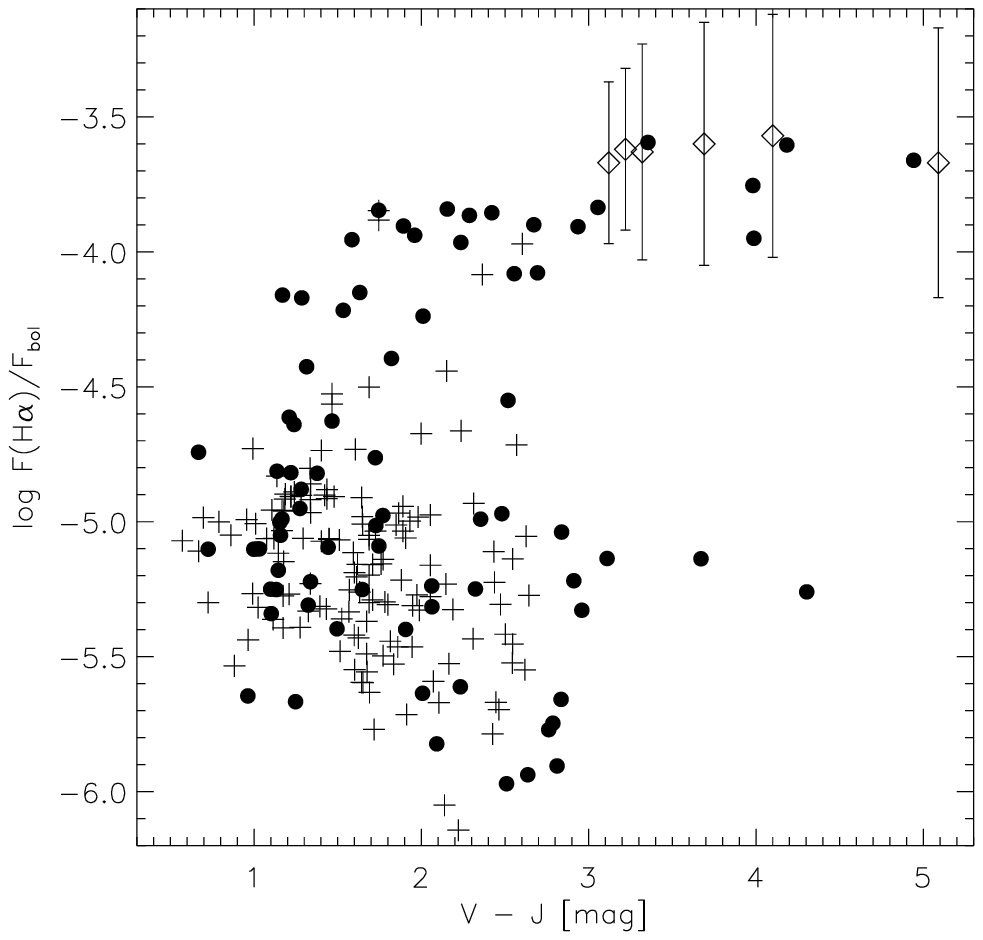}
   \hspace{0.2cm}
   \includegraphics[width=8.3cm]{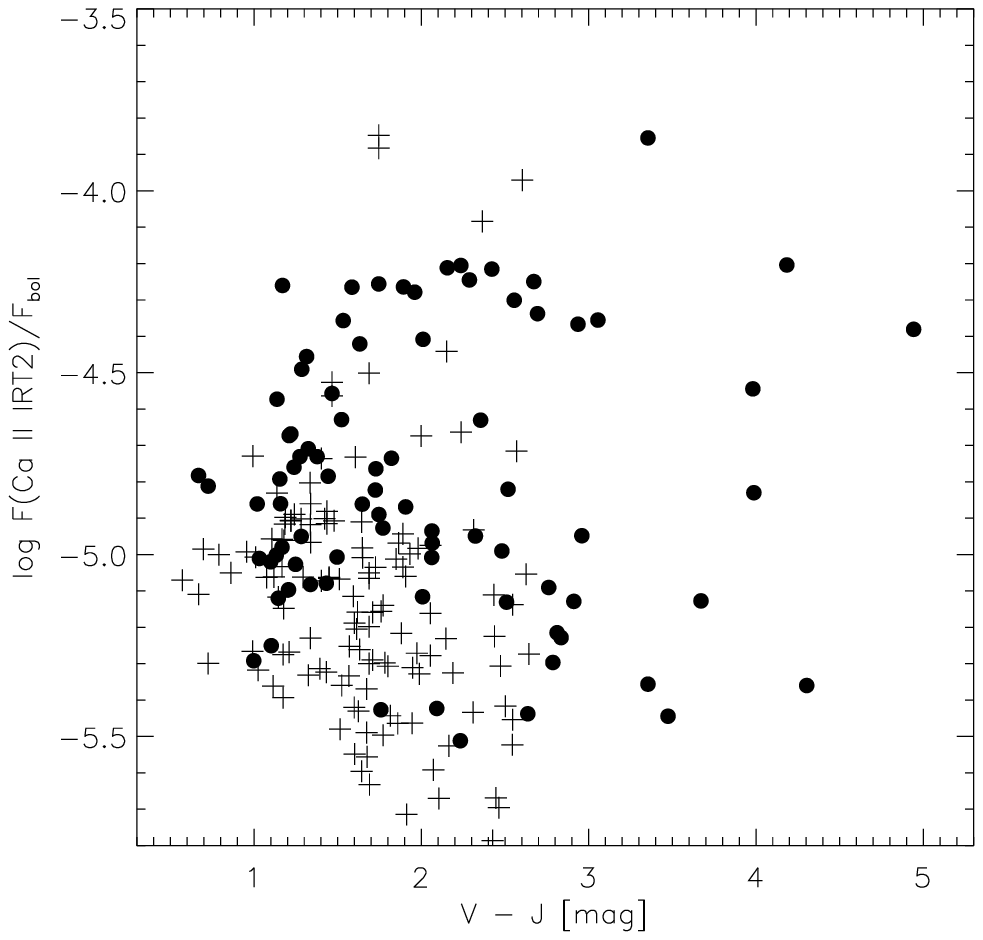}
   \caption{\textbf{Left:} $\log F_\mathrm{S}$(H$_{\alpha}$)/F$_\mathrm{bol}$ 
             versus $V-J$ for the stars in our sample (dots). Plusses are the 
             stars in \citet{mar09}. Diamonds are the data of \citet{wes04}.
             \textbf{Right:} $\log  F_\mathrm{S}$(\ion{Ca}{ii} $\lambda 8542$ \AA)/F$_\mathrm{bol}$  
             versus $V-J$ of the stars in our sample. Symbols are the same as in the left figure.}
   \label{f3}
\end{figure*}

In Fig.~\ref{f2}, we show the distribution of fluxes in different chromospheric 
lines {(to allow statistical comparisons with other works)}. The sample has 
a peak towards high chromospheric fluxes in chromospheric calcium. In H$\alpha$, 
{two peaks are observed}. {The peak at high fluxes in each line is a 
consequence of the selection method. The double peak in H$\alpha$ is 
presumably caused by the presence of two populations of H$\alpha$ 
emitters: saturated and non-saturated. To verify this hypothesis, we constructed
plots of the ratio $F_\mathrm{line}/F_\mathrm{bol}$ versus temperature for each 
chromospheric line. In Fig.~\ref{f3}, we show the result for H$\alpha$ and the 
\ion{Ca}{ii} $\lambda$ 8542 \AA\ (IRT2) line. For H$\alpha$, two well-defined 
branches are observed. A similar behavior has been observed for coronal 
sources \citep[e.g.][]{lop09} and attributed to X-ray emission saturation. 
Similarly, the stars in our sample show saturation in H$\alpha$ 
at a mean value $\log F_\mathrm{H\alpha}/F_\mathrm{bol} \sim -3.8$, as it has been 
found for early to mid--M stars \citep[e.g.][]{wal04,wes08}. The effect of saturation
is less marked in other indicators, such as the \ion{Ca}{ii} lines (see Fig.~\ref{f3},
right).
}

In addition to the absolute fluxes in the different chromospheric lines, we also
determined the $R'_\mathrm{HK}$ index (see Table~\ref{tabA4}), which is 
defined as the ratio of the emission from the chromosphere in the 
\ion{Ca}{ii} H \& K lines to the total bolometric emission of the star, i.e.: 

\begin{equation}
R'_\mathrm{HK} = \frac{F'_\mathrm{H} + F'_\mathrm{K}}{\sigma T^4}
\end{equation}

where $F'_\mathrm{H}$ and $F'_\mathrm{K}$ are the chromospheric fluxes 
in  the \ion{Ca}{ii} H and K lines, respectively. Effective temperatures were 
determined using empirical calibrations with the color index $B-V$ 
\citep[e.g.][]{gra05}. {Such calibrations are valid for $B-V \le 1.5$. 
Only seven stars of our sample (three of them being giants) have values 
of $B-V$ above 1.5. For them, we extrapolated the color-temperature 
relation. For $B-V > 1.5$, the spread in temperatures is large 
\citep[see Fig.~14.6 in][]{gra05}. In general, for M dwarfs, differences 
between the value given by the relation and that obtained with other 
methods \citep[e.g.][]{flo96} of up to 100-200~K are observed. 
}

Figures~\ref{figA1}, \ref{figA2}, and \ref{figA3} show the spectra of the stars 
in the sample in spectral regions containing the \ion{Ca}{ii} K line, H$\alpha$, 
and part of the \ion{Ca}{ii} infrared triplet.

\subsection{Summary}

Radial and rotational velocities were derived from each observation of star 
in our sample using the cross-correlation technique. We derived mean radial 
and rotational velocities for each star. In some cases (see Sect.~3.1) we observed
large variations in the radial velocity of the star that we attributed to binarity. 
Mean values of radial velocities were used to derive space motions. 
We also determined spectral types from the spectra of stars. Table~\ref{tabA2}
summarizes these results.

With regard to spectral lines, we determined equivalent widths of all optical 
chromospheric activity indicators of each star, as well as the 
\ion{Li}{i} $\lambda$6708 \AA\ line (see Table~\ref{tabA2} for results of the \ion{Li}{i} 
line). To reveal chromospheric emission lines in the
spectra, we used the spectral subtraction technique. Non-active stars with spectral 
types similar to those of our targets were used as templates for the subtraction. Their
spectra were conveniently broadened and shifted to fit our 
targets (see Sect.~3.4). The equivalent widths of the emission lines were converted 
into flux using equivalent width--flux relations. 
Tables~\ref{tabA3} and \ref{tabA4} summarize the results of the spectroscopic survey. 

{For completeness, we performed a simple statistical study of our results on chromospheric 
activity indicators. In our sample, a large spread in fluxes is observed for the different activity indicators.
The spread is especially noticeable for the H$\alpha$ line (Fig.~\ref{f2}, middle). To investigate 
this finding, we analyzed the $F_\mathrm{H\alpha}/F_\mathrm{bol}$ and 
$F_\ion{Ca}{ii}/F_\mathrm{bol}$ ratios as a function of the color of the stars. 
The results indicate the presence of two populations of chromospherically active stars in our sample.
In the $\log F_\mathrm{H\alpha}/F_\mathrm{bol}$ -- ($V-J$) diagram (Fig.~\ref{f3}, left), two 
branches are clearly observed for $V -J \geq 1.4$ mag (corresponding to an early-K dwarf). 
This dichotomy 
is statistically significant. Out of the 79 stars in our sample with chromospheric H$\alpha$ emission, 
53 have $V - J > 1.4$: with 23 stars in the upper branch and the remaining 30
in the lower one. The two groups have mean values $\log F_\mathrm{H\alpha} = -3.9$ with 
variance 0.04 and $\log F_\mathrm{H\alpha} = -5.3$ with variance 0.13, respectively. 
A simple two-sample t-test \citep[e.g.][]{sne89} assures that the two means are different 
(i.e. the two branches are statistically different), with a significance of 0.01 (corresponding
to a probability of 99$\%$).
The presence of the two branches is less clear when using other chromospheric activity 
indicators (see Fig.~\ref{f3}, right). 
}

{In a recent work, \citet{lop09} demonstrated that the sample of possible members of 
the young stellar kinematic groups of \citet{mon01a} is partially \textit{contaminated} by active field 
stars that do not belong to the moving groups. In that work, the stars in the X-ray saturation 
regime also showed high H$\alpha$ fluxes. In their Fig.~4, \citet{lop09} observed two 
branches: one for the high H$\alpha$ emitters and the other for the remaining active stars. 
Those stars populating the upper branch were indeed young ($\sim 10-120$ Myr). In contrast, 
field stars populated the lower branch. The study of \citet{lop09} was performed 
using part of the data presented by us in this work. Therefore, their results on the 
nature of the two populations of chromospheric active stars are applicable here. In fact, the stars
in the top branch in Fig.~\ref{f3} (left) are known to be young stars \citep[see][]{lop06} and very 
active M dwarfs. A similar conclusion has already been reached by \citet{vau80} for G and K 
stars. In their study, the authors observed two branches in the $\log S$ -- ($B-V$) diagram
with a gap between them. Vaughan \& Preston called them the young, active branch
and the less active, old branch.
Recent results show that this dichotomy between very active stars and less active ones
is also present in M dwarfs. In Fig.~\ref{f3} (left), we have over-plotted the data obtained by 
\citet{wes04} for the active M stars in the SLOAN Digital Sky Survey (the dispersion 
observed by the authors for each sub-spectral type is represented by vertical bars). 
Although our method for measuring the line flux is slightly different from that of 
\citet{wes04}, the figure shows clearly that M active dwarfs are located over the
young active branch. A similar trend can be observed for the early M dwarfs of 
\citet{reiners07}. 
In general, these results suggest that H$\alpha$ saturates equally in K to M stars.  
}


\begin{acknowledgements}
      Part of this work was co-supported by the Spanish 
      \emph{MICINN, Ministerio de Ciencia e Innovaci\'on} project
      numbers: AYA2008-00695 and AYA2008-06423-C03-03.
      J. L\'opez-Santiago is a postdoctoral fellow of the
      ASTROCAM (Red de Astrof\'isica en la CAM) with project 
      number S-0505/ESP/000237 of the IV PRICIT
      \emph{Plan Regional de Investigaci\'on Cient\'ifica e Innovaci\'on 
      Tecnol\'ogica de la Comunidad de Madrid}. The authors would like 
      to thank the referee for useful comments and suggestions that 
      allowed us to improve our work. 
\end{acknowledgements}

\appendix

\section{Description of the on-line material}

\begin{itemize}

\item Table~\ref{tabA1}: spectral-type reference stars and radial velocity 
          standards used for the subtraction of the photospheric spectrum and
          for determining radial and rotational velocities. Column \#1 is the 
          name/identification of the star; column \#2 is the spectral type;
          column \#3 is the radial velocity (and standard deviation); 
          column \#4 is the bibliographic reference for the radial velocity; 
          column \#5 is the rotational velocity (and standard deviation); 
          column \#6 is the bibliographic reference of the rotational velocity;
          column \#7 is the activity index S (and standard deviation); and
          column \#8 is the observing run in which the star was observed 
          by us.
\\
\item Table~\ref{tabA2}: spectroscopic parameters of the stars in the sample. 
          Each line correspond to a measurement/observation of the star.
           Column \#1 is the HD number or other name of the star; column \#2
           is the observing run in which the observation was performed; 
           column \#3 is the Modified Julian Date of the observation; column \#4
           is the spectral type of the star determined by us; column \#5 is the 
           B-V color from Tycho-2; column \#6 is the radial velocity determined 
           by us in that observation; column \#7 is a mean radial velocity of 
           every observation performed for the star; column \#8, \#9, and \#10 are
           the Galactic velocities; column \#11 is the rotational velocity determined
           in that observation; column \#12 is a mean value of the rotational 
           velocity of the star determined in each observation; column \#13 is 
           the photometric period found in the literature; column \#14 is the 
           equivalent width of the lithium line in 6707.8 \AA\ determined in that
           observation; column \#15 is a mean value of the equivalent width of the 
           lithium line of the star; and column \#16  is the preliminary assignation 
           to a moving group made by us in base to the Galactic velocity given in 
           columns \#8, \#9, and \#10.
\\
\item Table~\ref{tabA3}: equivalent widths of the different chromospheric lines 
          determined in each observation from the subtracted spectrum. 
           Column \#1 is the identification of the star in our sample (the same as
           in Table~\ref{tabA2}; column \#2 is the observing run of the observation;
           column \#3 is the Modified Julian Date of the observation; columns \#4 
           to \#16 are the equivalent widths in the chromospheric lines (in order):
           \ion{Ca}{ii} K \& H, H$\epsilon$, H$\delta$, H$\gamma$, H$\beta$, 
           \ion{He}{i} D$_3$, \ion{Na}{i} D$_2$ \& D$_1$, H$\alpha$, and \ion{Ca}{ii} 
           infrared triplet ($\lambda\lambda$ 8498, 8542, 8662 \AA), with errors.
\\
\item Table~\ref{tabA4}: Surface fluxes of the different chromospheric lines 
          determined in each observation from the equivalent widths in 
          Table~\ref{tabA3}. Column \#1 is the identification of the star in our sample 
          (the same as in Table~\ref{tabA2}; column \#2 is the observing run of the 
          observation; column \#3 is the Modified Julian Date of the observation; 
          columns \#4 
          to \#16 are the surface fluxes in the chromospheric lines (in order):
          \ion{Ca}{ii} K \& H, H$\epsilon$, H$\delta$, H$\gamma$, H$\beta$, 
          \ion{He}{i} D$_3$, \ion{Na}{i} D$_2$ \& D$_1$, H$\alpha$, and \ion{Ca}{ii} 
          infrared triplet ($\lambda\lambda$8498, 8542, 8662 \AA), with errors;
          column \#17 is the $R'_\mathrm{HK}$ index.          
\\
\item Figures~\ref{figA1} to \ref{figA4}: figures with the normalized spectrum 
         of the stars in our sample in the spectral regions of the \ion{Ca}{ii} K, 
         H$\alpha$, \ion{Li}{i}, and \ion{Ca}{ii} $\lambda\lambda$8498 and 
         8542 \AA\ lines, respectively.

\end{itemize}

\clearpage
\onllongtab{1}{
\begin{table*}
\caption[]{Spectral-type reference stars and radial velocity standards (marked with *).
\label{tabA1}}
\scriptsize
\begin{tabular}{llcccccl}
\noalign{\smallskip} \hline \hline \noalign{\smallskip}
HD/GJ & SpT & $V_{\rm r} \pm \sigma_{\rm V_r}$  & Ref$_{\rm v}$ & $v \sin i$ & 
Ref$_{\rm r}$ & $S \pm \sigma_{\rm S}$ & Observing run \\
   &              &          (km s$^{-1}$)            &             & (km s$^{-1}$) &
               &   &           \\
\noalign{\smallskip} \hline \noalign{\smallskip}
212754   & F7 V   & -17.8 $\pm$ 1.2 & a & 7.9 $\pm$ 0.7 & j & 0.142 $\pm$ 0.001 & 4, 5 \\
43587  * & F9 V   & ~~4.6 $\pm$ 0.1 & b & ...           &   & 0.158 $\pm$ 0.001 & 5, 6 \\
84737  * & G0.5 V & ~~6.0 $\pm$ 1.1 & b & 2.8 $\pm$ 0.8 & j & 0.144 $\pm$ 0.000 & 10 \\
10307    & G2 V   & ...             &   & 2.1 $\pm$ 0.5 & k & 0.152 $\pm$ 0.003 & 7 \\
193664   & G3 V   & ~-4.7 $\pm$ 1.2 & a & ...           &   & 0.161 $\pm$ 0.004 & 7 \\
25680    & G5 V   & ~24.0 $\pm$ 0.1 & c & 7.0 $\pm$ 0.7 & j & 0.281 $\pm$ 0.000 & 5, 7 \\
31966    & G5 V   & -18.1 $\pm$ 0.1 & c & ...           &   & ...               & 5 \\
71148  * & G5 V   & -31.0 $\pm$ 0.7 & b & ...           &   & 1.570 $\pm$ 0.000 & 10 \\
159222 * & G5 V   & -50.5 $\pm$ 1.2 & b & ...           &   & 0.164 $\pm$ 0.002 & 1, 4, 11 \\
182488 * & G8 V   & -21.5           & d & 0.6 $\pm$ 0.5 & k & 0.155 $\pm$ 0.008 & 1, 4, 11 \\
48432    & K0 III & ~17.9 $\pm$ 0.2 & e & $<1.0$        & e & 0.120 $\pm$ 0.000 & 5, 6 \\
62509  * & K0 III & ~~3.2 $\pm$ 0.3 & b & 1.7 $\pm$ 0.5 & k & 0.140 $\pm$ 0.019 & 3 \\
100696 * & K0 III & ~~0.2 $\pm$ 0.5 & b & 1.2 $\pm$ 1.0 & l & ...               & 10 \\
197989   & K0 III & -10.6 $\pm$ 0.5 & a & 2.0 $\pm$ 0.5 & k & 0.104 $\pm$ 0.001 & 1, 2, 4, 7, 8, 11 \\
3651   * & K0 V   & -32.8 $\pm$ 0.8 & b & 2.2 $\pm$ 0.5 & k & 0.191 $\pm$ 0.001 & 2, 3, 4, 5, 7 \\
97004    & K0 V   & ~~5.4 $\pm$ 0.1 & c & ...           &   & ...               & 10 \\
112758   & K0 V   & ~-4.1 $\pm$ 1.2 & a & ...           &   & 0.206 $\pm$ 0.000 & 10 \\
136442 * & K0 V   & -45.6 $\pm$ 0.8 & b & ...           &   & ...               & 10 \\
185144   & K0 V   & ~26.7 $\pm$ 0.1 & c & 0.6 $\pm$ 0.5 & k & 0.195 $\pm$ 0.003 & 11 \\
201651   & K0 V   & -13.7 $\pm$ 1.2 & a & ...           &   & ...               & 4, 7, 8 \\
92588  * & K1 IV  & ~43.5 $\pm$ 0.3 & f & $<1.0$        & e & ...               & 9, 10 \\
10476  * & K1 V   & -33.9 $\pm$ 0.9 & b & 0.6 $\pm$ 0.5 & k & 0.192 $\pm$ 0.001 & 1, 5 \\
12929  * & K2 III & -14.6 $\pm$ 0.2 & g & 1.8 $\pm$ 0.5 & k & 0.118 $\pm$ 0.002 & 5 \\
124897 * & K2 III & ~-5.3 $\pm$ 0.3 & g & 3.3 $\pm$ 0.5 & k & 0.144 $\pm$ 0.012 & 6, 10 \\
161096 * & K2 III & -12.5 $\pm$ 0.3 & g & 2.5 $\pm$ 0.5 & k & 0.103 $\pm$ 0.002 & 1, 4, 11 \\
201196   & K2 IV  & -34.8 $\pm$ 0.2 & e & $<1.0$        & e & ...               & 1, 5 \\
4628   * & K2 V   & -10.1 $\pm$ 0.4 & f & 0.0 $\pm$ 0.5 & h & 0.223 $\pm$ 0.001 & 2, 5, 8, 12 \\
136713   & K2 V   & ~-6.0 $\pm$ 0.1 & c & 3.8 $\pm$ 5.7 & m & ...               & 6 \\
166620   & K2 V   & ~~6.9 $\pm$ 0.1 & h & 0.0 $\pm$ 0.4 & h & 0.193 $\pm$ 0.001 & 1, 2, 4, 6, 7, 9, 10, 11 \\
16160    & K3 V   & ~25.8 $\pm$ 0.1 & c & 1.0 $\pm$ 1.0 & h & 0.221 $\pm$ 0.002 & 8 \\
219134   & K3/4 V & -18.6 $\pm$ 0.1 & c & 2.1 $\pm$ 0.5 & k & 0.229 $\pm$ 0.003 & 1, 2, 4, 5, 11 \\
29139  * & K5 III & ~54.2 $\pm$ 0.2 & g & 2.0 $\pm$ 1.0 & e & ...               & 4, 6, 7 \\
154363   & K5 V   & ~34.1 $\pm$ 0.1 & c & 3.7 $\pm$ 5.9 &   & 0.197 $\pm$ 0.001 & 1, 6, 10 \\
201091   & K5 V   & ~~7.0 $\pm$ 0.1 & h & 0.0 $\pm$ 0.8 & h & 0.613 $\pm$ 0.006 & 1, 4, 5, 6, 7, 8, 10, 11 \\
GJ 910   & K5 V   & ~~2.0           & i & 0.0 $\pm$ 0.0 & m & ...               & 7 \\
151877   & K7 V   & ~~2.0 $\pm$ 0.1 & c & 0.0 $\pm$ 0.0 & m & ...               & 1 \\
201092   & K7 V   & ~~7.2 $\pm$ 0.1 & h & 1.7 $\pm$ 0.6 & h & 0.922 $\pm$ 0.011 & 1, 4, 5, 6, 7, 8, 10, 11 \\
GJ 466   & M0 V   & ~-5.0 $\pm$ 5.0 & a & ...           &   & ...               & 10 \\
147379   & M0 V   & -18.8 $\pm$ 0.1 & c & 4.2 $\pm$ 6.2 & m & 1.761 $\pm$ 0.160 & 11 \\
GJ 720A  & M0 V   & -25.0 $\pm$ 2.5 & a & 6.3 $\pm$ 1.7 & m & ...               & 7 \\
GJ 16    & M0/1 V & ...             &   & ...           &   & ...               & 8 \\
GJ 806   & M1.5 V & -24.7 $\pm$ 0.1 & c & 1.9 $\pm$ 0.7 & n & ...               & 11 \\
18884  * & M2 III & -26.1 $\pm$ 0.3 & g & ...           &   & 0.331 $\pm$ 0.004 & 8 \\
115521 * & M2 III & -28.6 $\pm$ 2.3 & g & ...           &   & ...               & 6, 10 \\
95735    & M2 V   & -84.7 $\pm$ 0.1 & c & 0.0 $\pm$ 0.0 & n & 0.392 $\pm$ 0.009 & 6, 10 \\
GJ 687B$^\dag$ & M3.5 V & -28.8 $\pm$ 0.1 & c & ...     &   & ...               & 6, 10 \\
\noalign{\smallskip} \hline \noalign{\smallskip}
\end{tabular}

\scriptsize
\dag \ The radial velocity given for GJ~687B is that of GJ~687A.

a \citet{duf95}, WEB (Wilson Evans Batten Catalogue).

b \citet{bar86}. 

c \citet{nid02}. 

d {\it ELODIE}.\index{ELODIE}

e \citet{dem99}. 

f \citet{bea79}. 

g \citet{udr99}. 

h \citet{ben84}. 

i \citet{dye54}. 

j \citet{sod82}, \citet{sod89}. 

k \citet{fek97}. 

l \citet{dem00}. 

m \citet{tok92}. 

n \citet{mar92}. 
\end{table*}
}

\clearpage
\onllongtab{2}{\tiny
\begin{landscape}
\begin{longtable}{lccccrrrrrrrrrrc}
\caption{Spectroscopic results of the sample. Those stars marked with $^*$ in column \#16 (MG) 
are inside the boundaries of the young disk population but have large $W$ velocities. 
\label{tabA2}}\\
\noalign{\smallskip}
\hline \hline
\noalign{\smallskip}
HD/ & ID & MJD & Sp.T. & $B-V$ & 
\multicolumn{1}{c}{$V_{\rm r}$} &
\multicolumn{1}{c}{$\overline{V}_{\rm r}$} &
\multicolumn{1}{c}{$U$} &
\multicolumn{1}{c}{$V$} &
\multicolumn{1}{c}{$W$} &
\multicolumn{1}{c}{$v \sin i$} &
\multicolumn{1}{c}{$\overline{v \sin i}$} &
\multicolumn{1}{c}{$P_{\rm phot}$} &
\multicolumn{1}{c}{$EW({\rm Li I})$} &
\multicolumn{1}{c}{$\overline{EW({\rm Li I})}$} &
\multicolumn{1}{c}{MG} \\
\noalign{\smallskip}
 other name & & (days) &  & (mag) & 
\multicolumn{1}{c}{(km s$^{-1}$)} &
\multicolumn{1}{c}{(km s$^{-1}$)} &
\multicolumn{1}{c}{(km s$^{-1}$)} &
\multicolumn{1}{c}{(km s$^{-1}$)} &
\multicolumn{1}{c}{(km s$^{-1}$)} &
\multicolumn{1}{c}{(km s$^{-1}$)} &
\multicolumn{1}{c}{(km s$^{-1}$)} &
\multicolumn{1}{c}{(days)} &
\multicolumn{1}{c}{(m\AA)} &
\multicolumn{1}{c}{(m\AA)} \\
\noalign{\smallskip}
\hline
\noalign{\smallskip}
\endfirsthead
\caption{Continued.}\\
\hline
\noalign{\smallskip}
HD/ & ID & MJD & Sp.T. & $B-V$ &
\multicolumn{1}{c}{$V_{\rm r}$} &
\multicolumn{1}{c}{$\overline{V}_{\rm r}$} &
\multicolumn{1}{c}{$U$} &
\multicolumn{1}{c}{$V$} &
\multicolumn{1}{c}{$W$} &
\multicolumn{1}{c}{$v \sin i$} &
\multicolumn{1}{c}{$\overline{v \sin i}$} &
\multicolumn{1}{c}{$P_{\rm phot}$} &
\multicolumn{1}{c}{$EW({\rm Li I})$} &
\multicolumn{1}{c}{$\overline{EW({\rm Li I})}$} &
\multicolumn{1}{c}{MG} \\
\noalign{\smallskip}
 other name & & (days) &  & (mag) & 
\multicolumn{1}{c}{(km s$^{-1}$)} &
\multicolumn{1}{c}{(km s$^{-1}$)} &
\multicolumn{1}{c}{(km s$^{-1}$)} &
\multicolumn{1}{c}{(km s$^{-1}$)} &
\multicolumn{1}{c}{(km s$^{-1}$)} &
\multicolumn{1}{c}{(km s$^{-1}$)} &
\multicolumn{1}{c}{(km s$^{-1}$)} &
\multicolumn{1}{c}{(days)} &
\multicolumn{1}{c}{(m\AA)} &
\multicolumn{1}{c}{(m\AA)} \\
\noalign{\smallskip}
\hline
\noalign{\smallskip}
\endhead
\hline
\endfoot
\hline
\endlastfoot
            166 & (5)  &  51855.026 & K0     V     & 0.75 &               ... &    -6.9 $\pm$ 0.1 & -15.00 & -21.60 & -10.04 &  6.54 $\pm$ 1.00 &  6.54 $\pm$ 1.00 &  5.69   &  78.0 $\pm$ 0.5 &  75.5 $\pm$ 0.3  & LA \\
                & (5)  &  51856.940 &              &      &  -6.94 $\pm$ 0.10 &                   &        &        &        &  6.54 $\pm$ 1.00 &                  &      &  73.0 $\pm$ 0.5 &                 \\
           1405 & (1)  &  51384.173 & K2     V     & 1.04 & -11.76 $\pm$ 0.59 &   -11.2 $\pm$ 0.1 &  -5.34 & -28.85 & -17.84 & 23.12 $\pm$ 1.60 & 23.88 $\pm$ 0.29 &  1.74   & 263.4 $\pm$ 1.7 & 268.2 $\pm$ 0.3 & LA \\
                & (1)  &  51385.040 &              &      & -11.49 $\pm$ 0.23 &                   &        &        &        & 20.47 $\pm$ 2.07 &                  &      & 256.4 $\pm$ 1.7 &                 \\
                & (1)  &  51386.101 &              &      & -12.50 $\pm$ 0.37 &                   &        &        &        & 22.01 $\pm$ 1.39 &                  &      & 265.4 $\pm$ 1.7 &                 \\
                & (1)  &  51387.039 &              &      & -10.51 $\pm$ 0.42 &                   &        &        &        & 22.51 $\pm$ 1.47 &                  &      & 263.4 $\pm$ 1.7 &                 \\
                & (1)  &  51388.125 &              &      & -11.95 $\pm$ 0.26 &                   &        &        &        & 22.01 $\pm$ 1.32 &                  &      & 250.4 $\pm$ 1.7 &                 \\
                & (1)  &  51389.081 &              &      & -11.11 $\pm$ 0.47 &                   &        &        &        & 22.96 $\pm$ 1.43 &                  &      & 269.4 $\pm$ 1.7 &                 \\
\end{longtable}
\end{landscape}
}

\clearpage
\onllongtab{3}{\scriptsize
\begin{landscape}
\begin{longtable}{lccccccccccccccc}
\caption{Chromospheric emission lines equivalent widths
\label{tabA3}}\\
\noalign{\smallskip}
\hline \hline
\noalign{\smallskip}
 & & & \multicolumn{13}{c}{$EW$(\AA) in the subtracted spectrum} \\
\cline{4-16}
\noalign{\smallskip}
HD/ & ID & MJD &
\multicolumn{2}{c}{Ca {\sc ii}} & & & & & He {\sc i} &
\multicolumn{2}{c}{Na {\sc i}} & &
\multicolumn{3}{c}{Ca {\sc ii} IRT} \\
\cline{4-5}\cline{11-12}\cline{14-16}
\noalign{\smallskip}
other name & & (days) & K & H &
H$\epsilon$ & H$\delta$ & H$\gamma$ & H$\beta$ &
D$_3$ & D$_2$ & D$_1$ & H$\alpha$ &
$\lambda$8498 & $\lambda$8542 & $\lambda$8662 \\
\noalign{\smallskip}
\hline
\noalign{\smallskip}
\endfirsthead
\caption{continued.}\\
\hline \hline
\noalign{\smallskip}
 & & & \multicolumn{13}{c}{$EW$(\AA) in the subtracted spectrum} \\
\cline{4-16}
\noalign{\smallskip}
HD/ & ID & MJD &
\multicolumn{2}{c}{Ca {\sc ii}} & & & & & He {\sc i} &
\multicolumn{2}{c}{Na {\sc i}} & &
\multicolumn{3}{c}{Ca {\sc ii} IRT} \\
\cline{4-5}\cline{11-12}\cline{14-16}
\noalign{\smallskip}
other name & & (days) & K & H &
H$\epsilon$ & H$\delta$ & H$\gamma$ & H$\beta$ &
D$_3$ & D$_2$ & D$_1$ & H$\alpha$ &
$\lambda$8498 & $\lambda$8542 & $\lambda$8662 \\
\noalign{\smallskip}
\hline
\noalign{\smallskip}
\endhead
\hline
\endfoot
\hline
\endlastfoot
166               &  (5)   &  51855.0269   & 0.32 $\pm$ 0.10 & ... & ... & ... & ... & ... & ... & ... & ... & 0.06 $\pm$ 0.02 & 0.18 $\pm$ 0.10 & 0.22 $\pm$ 0.10 & ... \\
                  &  (5)   &  51856.9403   & 0.20 $\pm$ 0.04 & ... & ... & ... & ... & ... & ... & ... & ... & 0.05 $\pm$ 0.02 & 0.15 $\pm$ 0.09 & 0.17 $\pm$ 0.15 & ... \\
1405              &  (1)   &  51384.1732   & 1.78 $\pm$ 0.10 & 1.17 $\pm$ 0.05 & 0.31 $\pm$ 0.06 & 0.27 $\pm$ 0.04 & 0.29 $\pm$ 0.06 & 0.47 $\pm$ 0.04 & ... & ... & ... & 1.39 $\pm$ 0.16 & 0.49 $\pm$ 0.07 & 0.63 $\pm$ 0.08 & 0.52 $\pm$ 0.06 \\
                  &  (1)   &  51385.0401   & 1.36 $\pm$ 0.15 & 0.99 $\pm$ 0.06 & 0.33 $\pm$ 0.06 & 0.28 $\pm$ 0.06 & 0.28 $\pm$ 0.05 & 0.45 $\pm$ 0.04 & ... & ... & ... & 1.26 $\pm$ 0.08 & 0.51 $\pm$ 0.07 & 0.63 $\pm$ 0.08 & 0.56 $\pm$ 0.03 \\
                  &  (1)   &  51386.1011   & 1.98 $\pm$ 0.23 & 0.93 $\pm$ 0.06 & 0.32 $\pm$ 0.06 & 0.26 $\pm$ 0.06 & 0.29 $\pm$ 0.06 & 0.51 $\pm$ 0.05 & ... & ... & ... & 1.35 $\pm$ 0.07 & 0.50 $\pm$ 0.06 & 0.66 $\pm$ 0.07 & 0.53 $\pm$ 0.02 \\
                  &  (1)   &  51387.0396   & ... & ... & ... & 0.28 $\pm$ 0.09 & 0.27 $\pm$ 0.06 & 0.44 $\pm$ 0.05 & ... & ... & ... & 1.23 $\pm$ 0.11 & 0.47 $\pm$ 0.07 & 0.62 $\pm$ 0.08 & 0.55 $\pm$ 0.05 \\
                  &  (1)   &  51388.1258   & 1.97 $\pm$ 0.10 & 0.82 $\pm$ 0.02 & 0.26 $\pm$ 0.03 & 0.24 $\pm$ 0.03 & 0.26 $\pm$ 0.04 & 0.48 $\pm$ 0.03 & ... & ... & ... & 1.20 $\pm$ 0.08 & 0.47 $\pm$ 0.07 & 0.61 $\pm$ 0.08 & 0.52 $\pm$ 0.03 \\
                  &  (1)   &  51389.0818   & 2.18 $\pm$ 0.08 & 1.68 $\pm$ 0.09 & 0.33 $\pm$ 0.10 & ... & ... & 0.47 $\pm$ 0.04 & ... & ... & ... & 1.36 $\pm$ 0.14 & 0.49 $\pm$ 0.07 & 0.68 $\pm$ 0.08 & 0.53 $\pm$ 0.05 \\
\end{longtable}
\end{landscape}
}

\clearpage
\onllongtab{4}{\scriptsize
\begin{landscape}
\begin{longtable}{lccccccccccccccccccc}
\caption{Chromospheric emission lines surface fluxes and $R`_\mathrm{HK}$
\label{tabA4}}\\
\noalign{\smallskip}
\hline \hline
\noalign{\smallskip}
& & & \multicolumn{16}{c}{$\log F_{\rm S}$ [erg cm$^{-2}$ s$^{-1}$] in the subtracted spectrum} \\
\cline{4-16}
\noalign{\smallskip}
HD/ & ID & MJD &
\multicolumn{2}{c}{Ca {\sc ii}} & & & & & He {\sc i} &
\multicolumn{2}{c}{Na {\sc i}} & &
\multicolumn{3}{c}{Ca {\sc ii} IRT} \\
\cline{4-5}\cline{11-12}\cline{14-16}
\noalign{\smallskip}
other name & & (days) & K & H &
H$\epsilon$ & H$\delta$ & H$\gamma$ & H$\beta$ &
D$_3$ & D$_2$ & D$_1$ & H$\alpha$ &
$\lambda$8498 & $\lambda$8542 & $\lambda$8662 & 
$R'_\mathrm{HK}$ \\
\noalign{\smallskip}
\hline
\noalign{\smallskip}
\endfirsthead
\caption{continued.}\\
\hline \hline
\noalign{\smallskip}
& & & \multicolumn{16}{c}{$\log F_{\rm S}$ [erg cm$^{-2}$ s$^{-1}$] in the subtracted spectrum} \\
\cline{4-16}
\noalign{\smallskip}
HD/ & ID & MJD &
\multicolumn{2}{c}{Ca {\sc ii}} & & & & & He {\sc i} &
\multicolumn{2}{c}{Na {\sc i}} & &
\multicolumn{3}{c}{Ca {\sc ii} IRT} \\
\cline{4-5}\cline{11-12}\cline{14-16}
\noalign{\smallskip}
other name & & (days) & K & H &
H$\epsilon$ & H$\delta$ & H$\gamma$ & H$\beta$ &
D$_3$ & D$_2$ & D$_1$ & H$\alpha$ &
$\lambda$8498 & $\lambda$8542 & $\lambda$8662 & 
$R'_\mathrm{HK}$ \\
\noalign{\smallskip}
\hline
\noalign{\smallskip}
\endhead
\hline
\endfoot
\hline
\endlastfoot
166               &  (5)   &  51855.0269   & 6.23 $\pm$ 0.31 & ... & ... & ... & ... & ... & ... & ... & ... & 5.51 $\pm$ 0.33 & 5.48 $\pm$ 0.56 & 5.57 $\pm$ 0.45 & ... & ... \\
                  &  (5)   &  51856.9403   & 6.03 $\pm$ 0.20 & ... & ... & ... & ... & ... & ... & ... & ... & 5.43 $\pm$ 0.40 & 5.75 $\pm$ 0.60 & 5.80 $\pm$ 0.88 & ... & ... \\
1405              &  (1)   &  51384.1732   & 6.40 $\pm$ 0.06 & 6.21 $\pm$ 0.04 & 5.64 $\pm$ 0.19 & 5.59 $\pm$ 0.15 & 5.65 $\pm$ 0.21 & 5.91 $\pm$ 0.09 & ... & ... & ... & 6.56 $\pm$ 0.12 & 6.06 $\pm$ 0.14 & 6.17 $\pm$ 0.13 & 6.09 $\pm$ 0.12 & -3.87 \\
                  &  (1)   &  51385.0401   & 6.28 $\pm$ 0.11 & 6.14 $\pm$ 0.06 & 5.66 $\pm$ 0.18 & 5.61 $\pm$ 0.21 & 5.63 $\pm$ 0.18 & 5.89 $\pm$ 0.09 & ... & ... & ... & 6.51 $\pm$ 0.06 & 6.08 $\pm$ 0.14 & 6.17 $\pm$ 0.13 & 6.12 $\pm$ 0.05 & -3.97 \\
                  &  (1)   &  51386.1011   & 6.44 $\pm$ 0.12 & 6.11 $\pm$ 0.06 & 5.65 $\pm$ 0.19 & 5.58 $\pm$ 0.23 & 5.65 $\pm$ 0.21 & 5.95 $\pm$ 0.10 & ... & ... & ... & 6.54 $\pm$ 0.05 & 6.07 $\pm$ 0.12 & 6.19 $\pm$ 0.11 & 6.09 $\pm$ 0.04 & -3.88 \\
                  &  (1)   &  51387.0396   & ... & ... & ... & 5.61 $\pm$ 0.32 & 5.62 $\pm$ 0.22 & 5.88 $\pm$ 0.11 & ... & ... & ... & 6.50 $\pm$ 0.09 & 6.04 $\pm$ 0.15 & 6.16 $\pm$ 0.13 & 6.11 $\pm$ 0.09 & ... \\
                  &  (1)   &  51388.1258   & 6.44 $\pm$ 0.05 & 6.06 $\pm$ 0.02 & 5.56 $\pm$ 0.12 & 5.54 $\pm$ 0.12 & 5.60 $\pm$ 0.15 & 5.92 $\pm$ 0.06 & ... & ... & ... & 6.49 $\pm$ 0.07 & 6.04 $\pm$ 0.15 & 6.16 $\pm$ 0.13 & 6.09 $\pm$ 0.06 & -3.89 \\
                  &  (1)   &  51389.0818   & 6.48 $\pm$ 0.04 & 6.37 $\pm$ 0.05 & 5.66 $\pm$ 0.30 & ... & ... & 5.91 $\pm$ 0.09 & ... & ... & ... & 6.55 $\pm$ 0.10 & 6.06 $\pm$ 0.14 & 6.20 $\pm$ 0.12 & 6.09 $\pm$ 0.09 & -3.76 \\
\end{longtable}
\end{landscape}
}

\newpage

\clearpage
\onlfig{1}{
\begin{figure*}{H}
\includegraphics[]{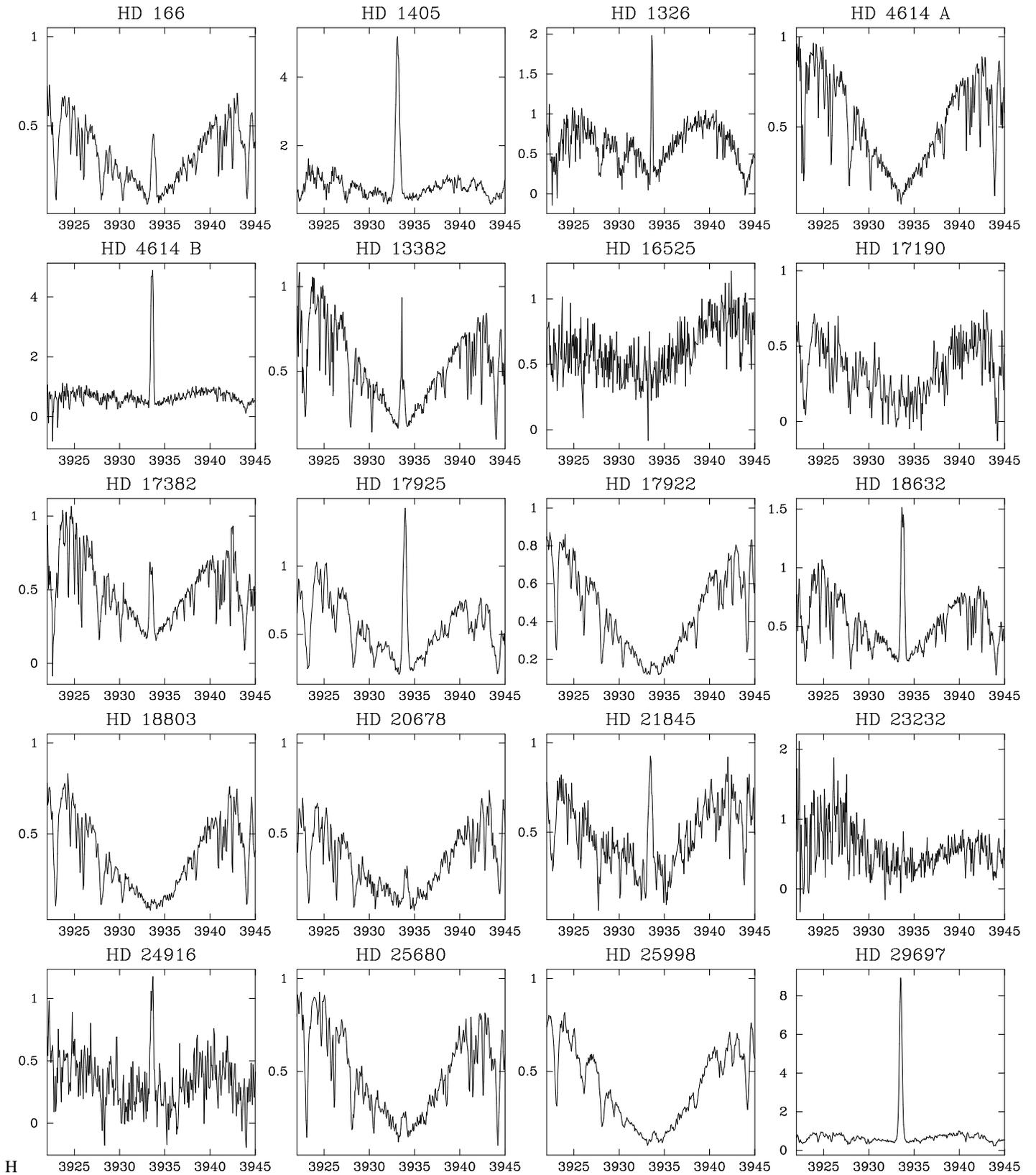}
\caption{\ion{Ca}{ii} K spectra of the stars of our sample with observations 
in the H \& K wavelength range.}
\label{figA1}
\end{figure*}
}
\clearpage
\onlfig{1}{
\begin{figure*}{H}
\includegraphics[]{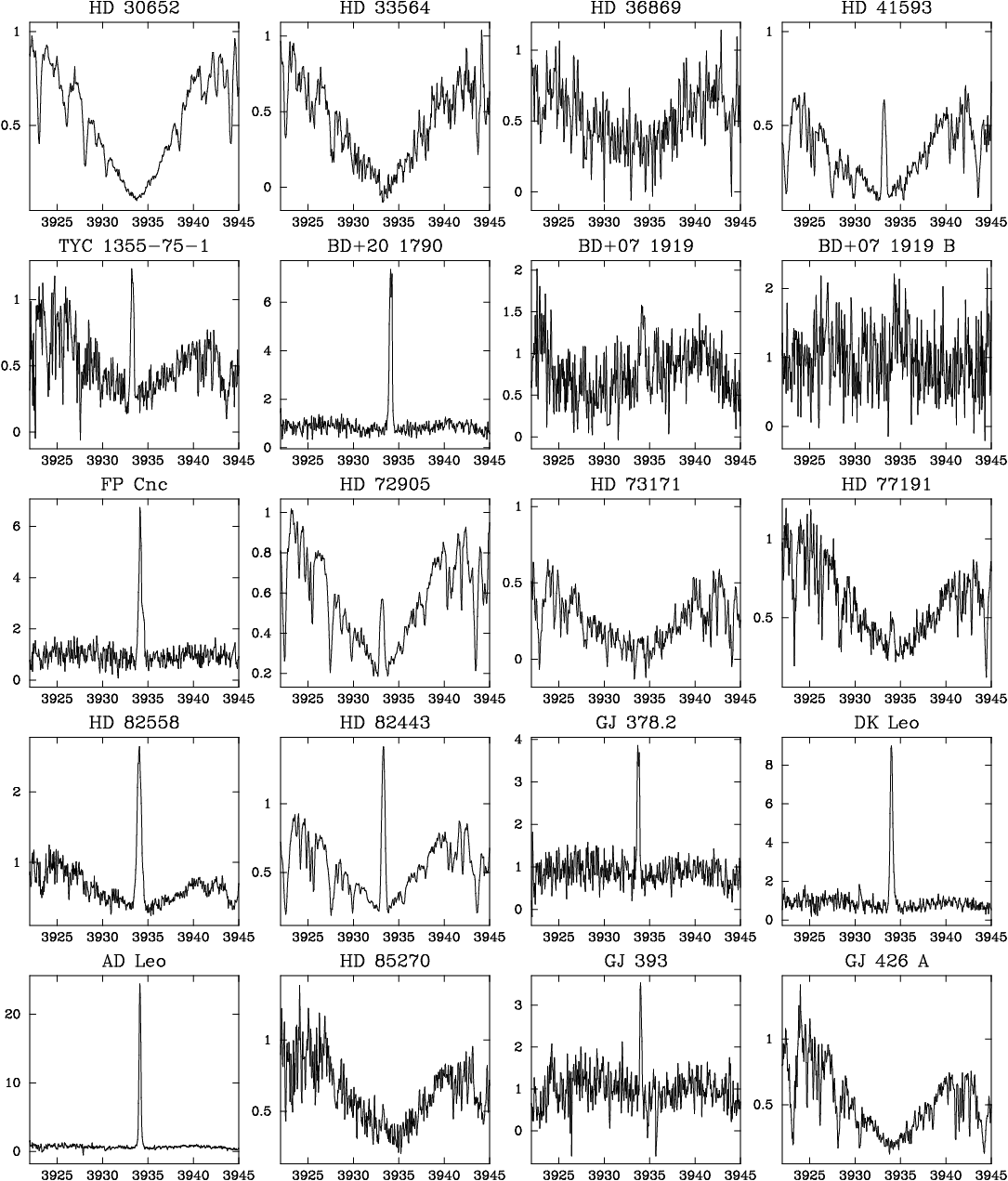}
\caption{Continued.}
\end{figure*}
}
\clearpage
\onlfig{1}{
\begin{figure*}{H}
\includegraphics[]{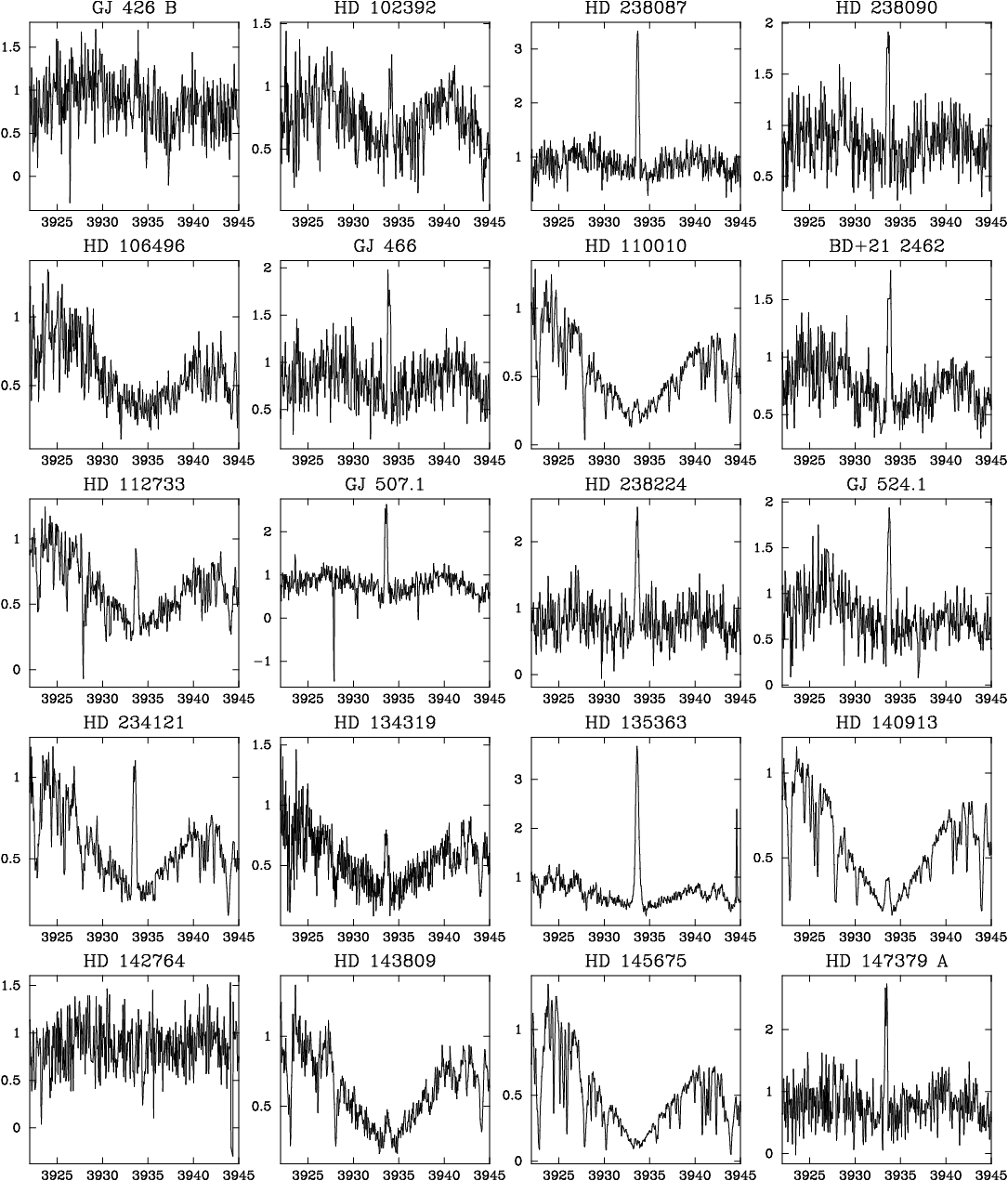}
\caption{Continued.}
\end{figure*}
}
\clearpage
\onlfig{1}{
\begin{figure*}{H}
\includegraphics[]{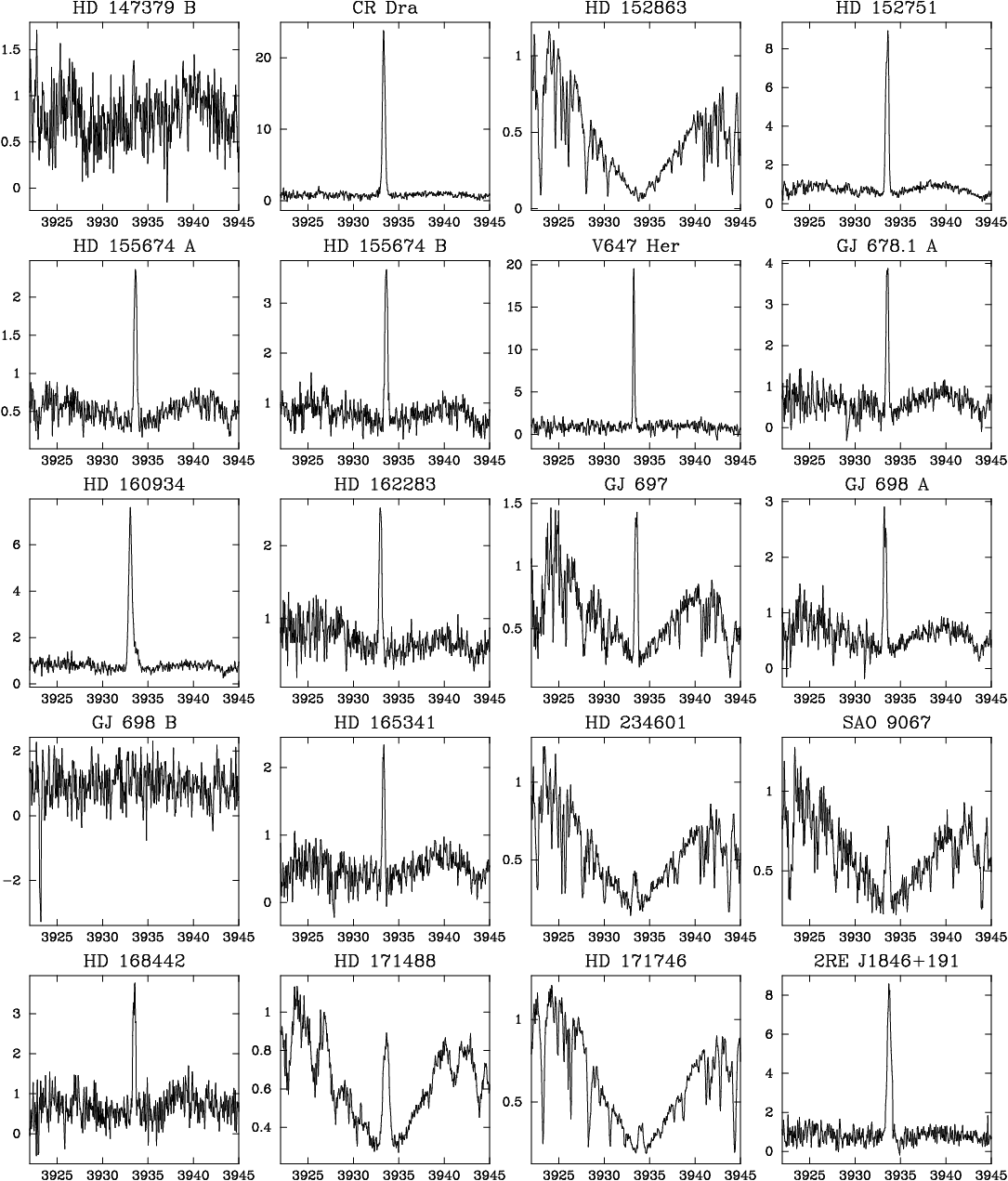}
\caption{Continued.}
\label{figA1}
\end{figure*}
}
\clearpage
\onlfig{1}{
\begin{figure*}{H}
\includegraphics[]{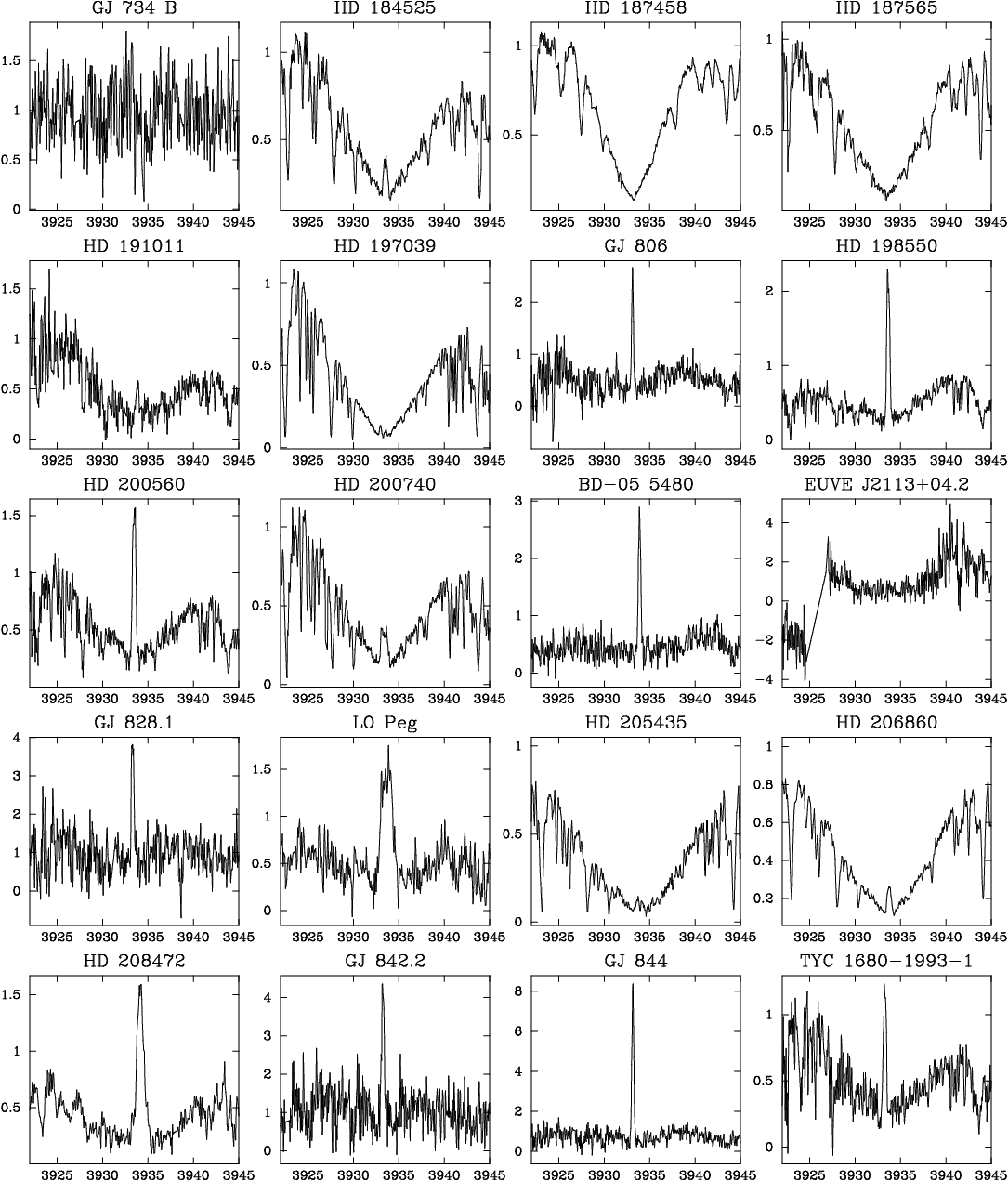}
\caption{Continued.}
\end{figure*}
}
\clearpage
\onlfig{1}{
\begin{figure*}{H}
\includegraphics[]{survey_cak_6.ps}
\caption{Continued.}
\end{figure*}
}

\clearpage
\onlfig{2}{
\begin{figure*}{H}
\includegraphics[]{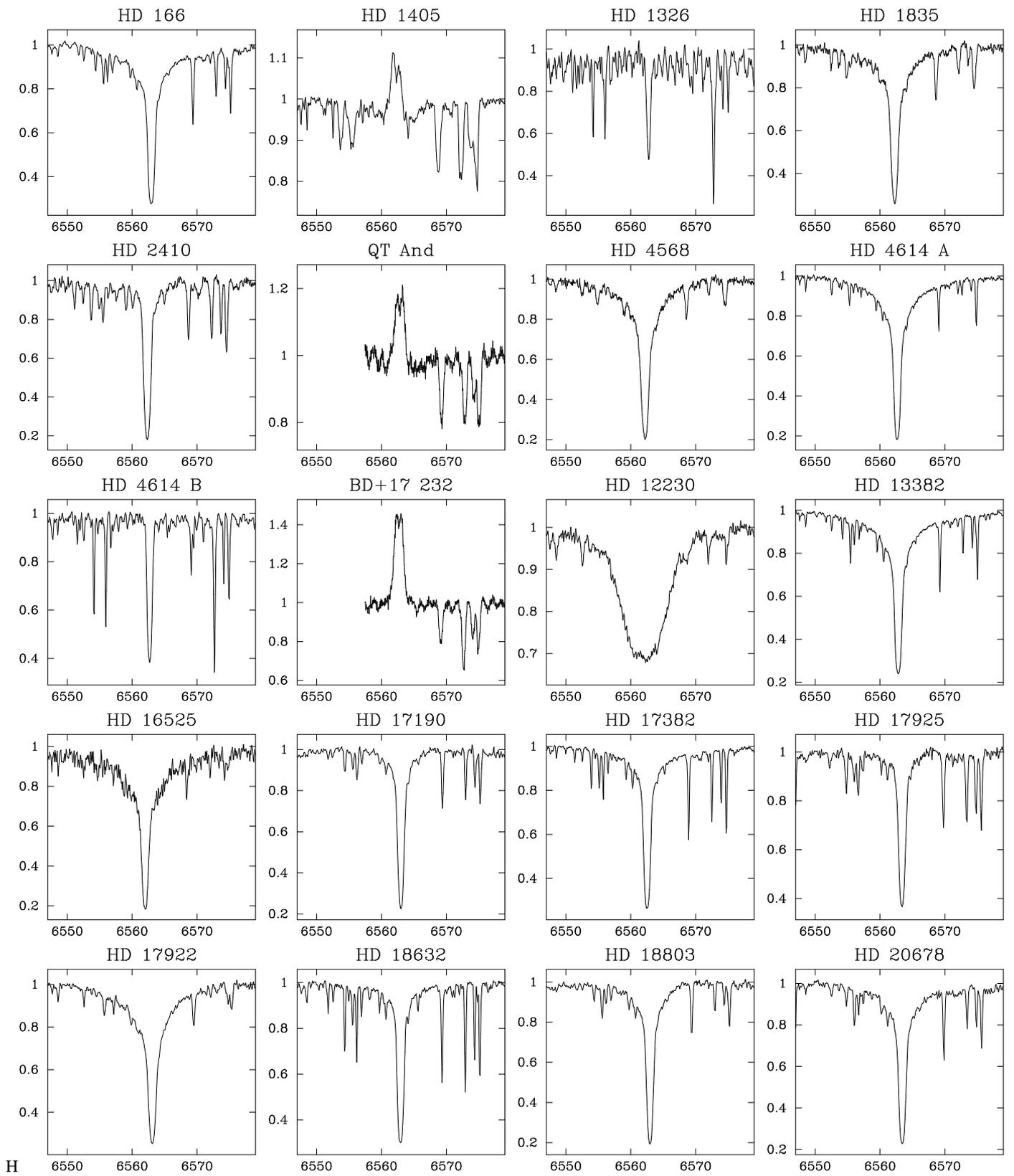}
\caption{H$\alpha$ spectra of the stars of our sample with observations 
in H$\alpha$.}
\label{figA2}
\end{figure*}
}
\clearpage
\onlfig{2}{
\begin{figure*}{H}
\includegraphics[]{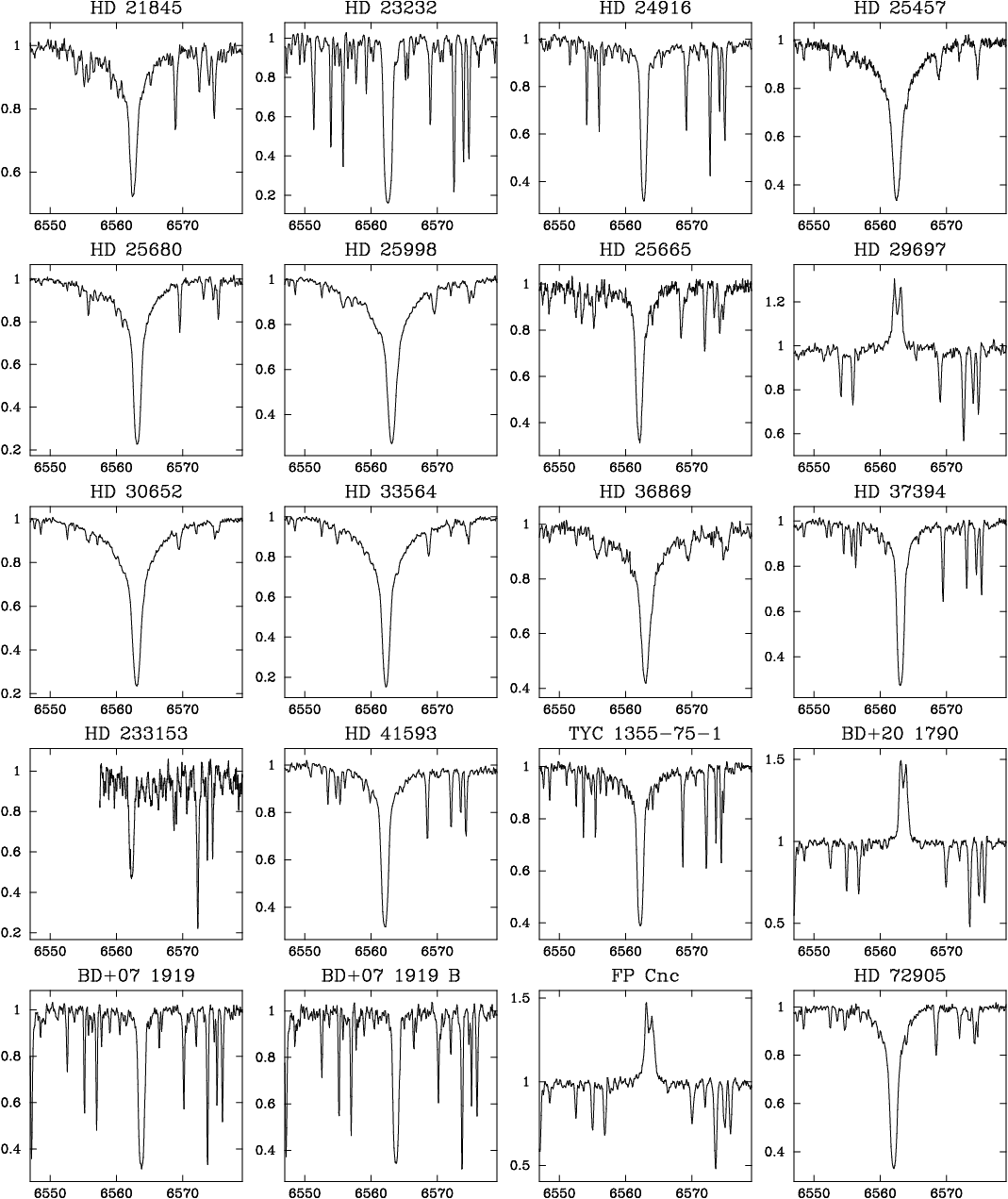}
\caption{Continued.}
\end{figure*}
}
\clearpage
\onlfig{2}{
\begin{figure*}{H}
\includegraphics[]{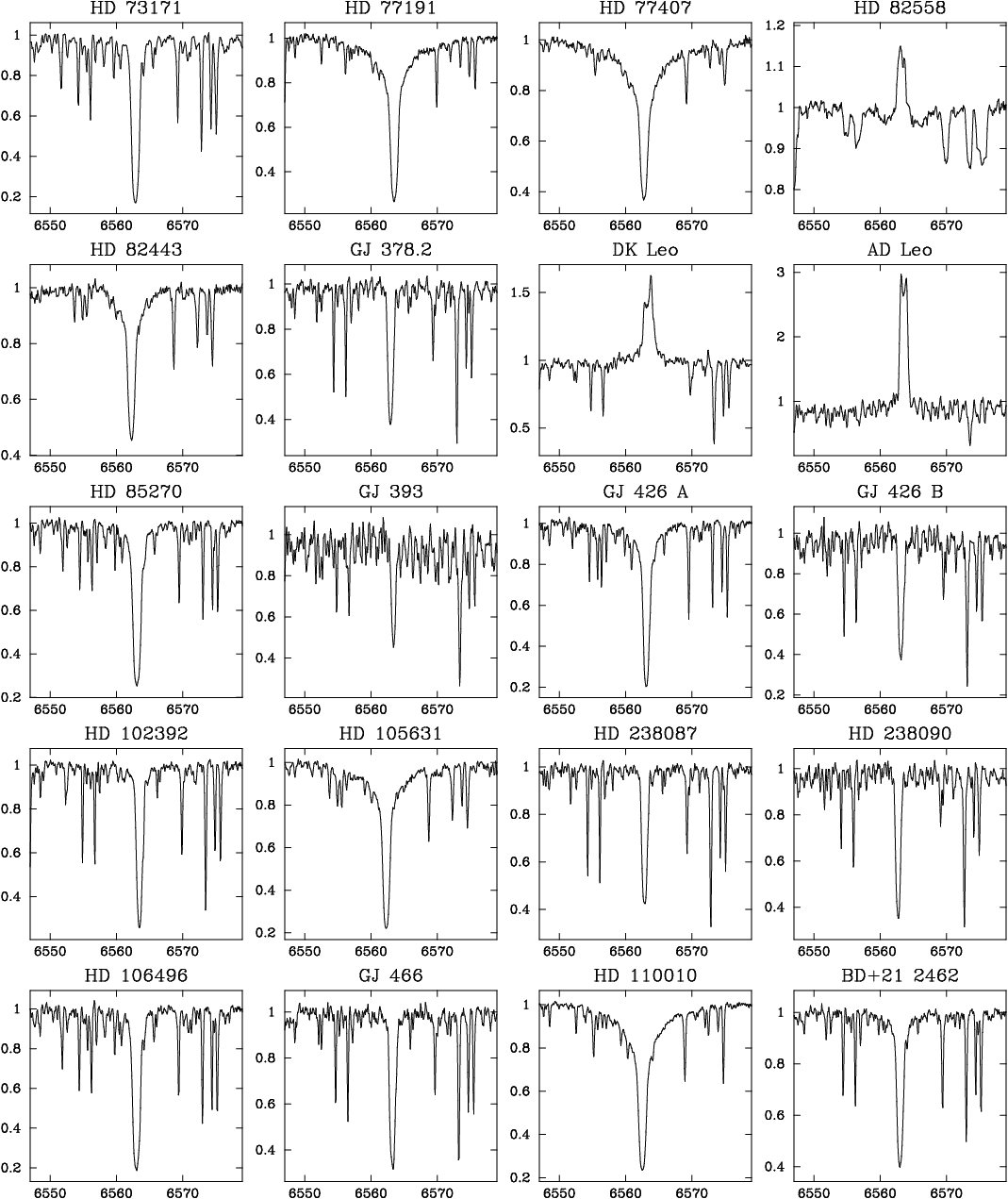}
\caption{Continued.}
\end{figure*}
}
\clearpage
\onlfig{2}{
\begin{figure*}{H}
\includegraphics[]{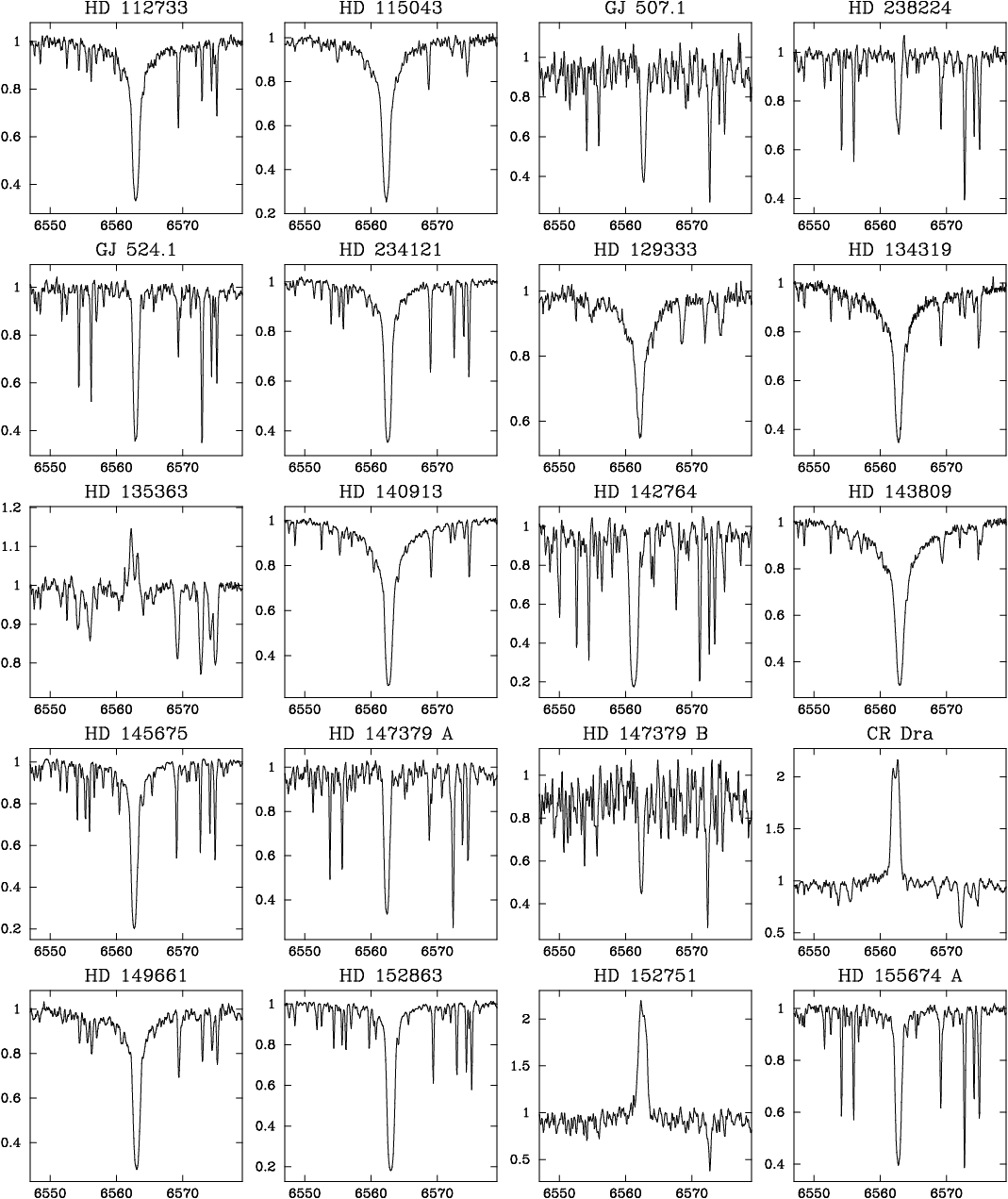}
\caption{Continued.}
\end{figure*}
}
\clearpage
\onlfig{2}{
\begin{figure*}{H}
\includegraphics[]{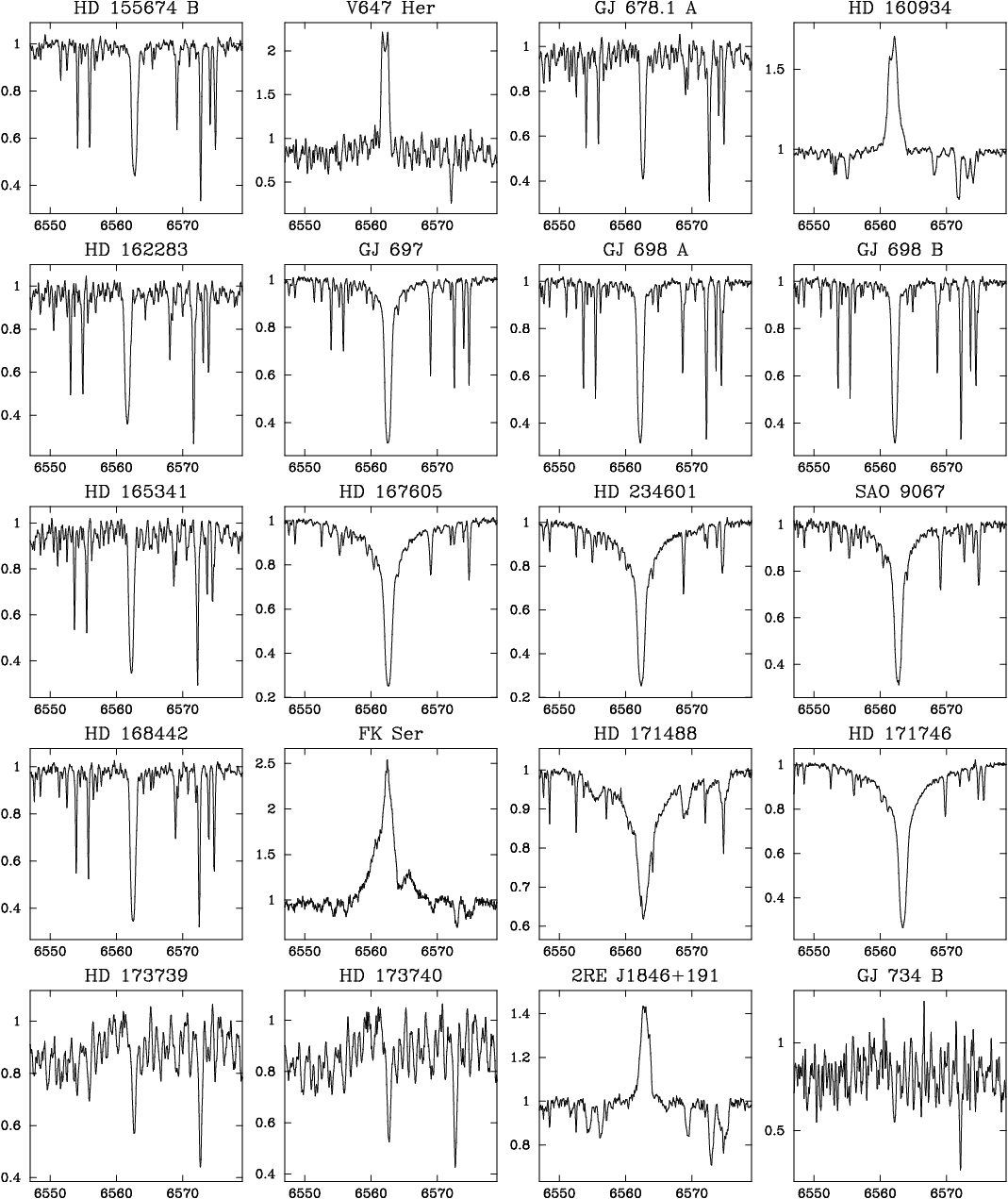}
\caption{Continued.}
\end{figure*}
}
\clearpage
\onlfig{2}{
\begin{figure*}{H}
\includegraphics[]{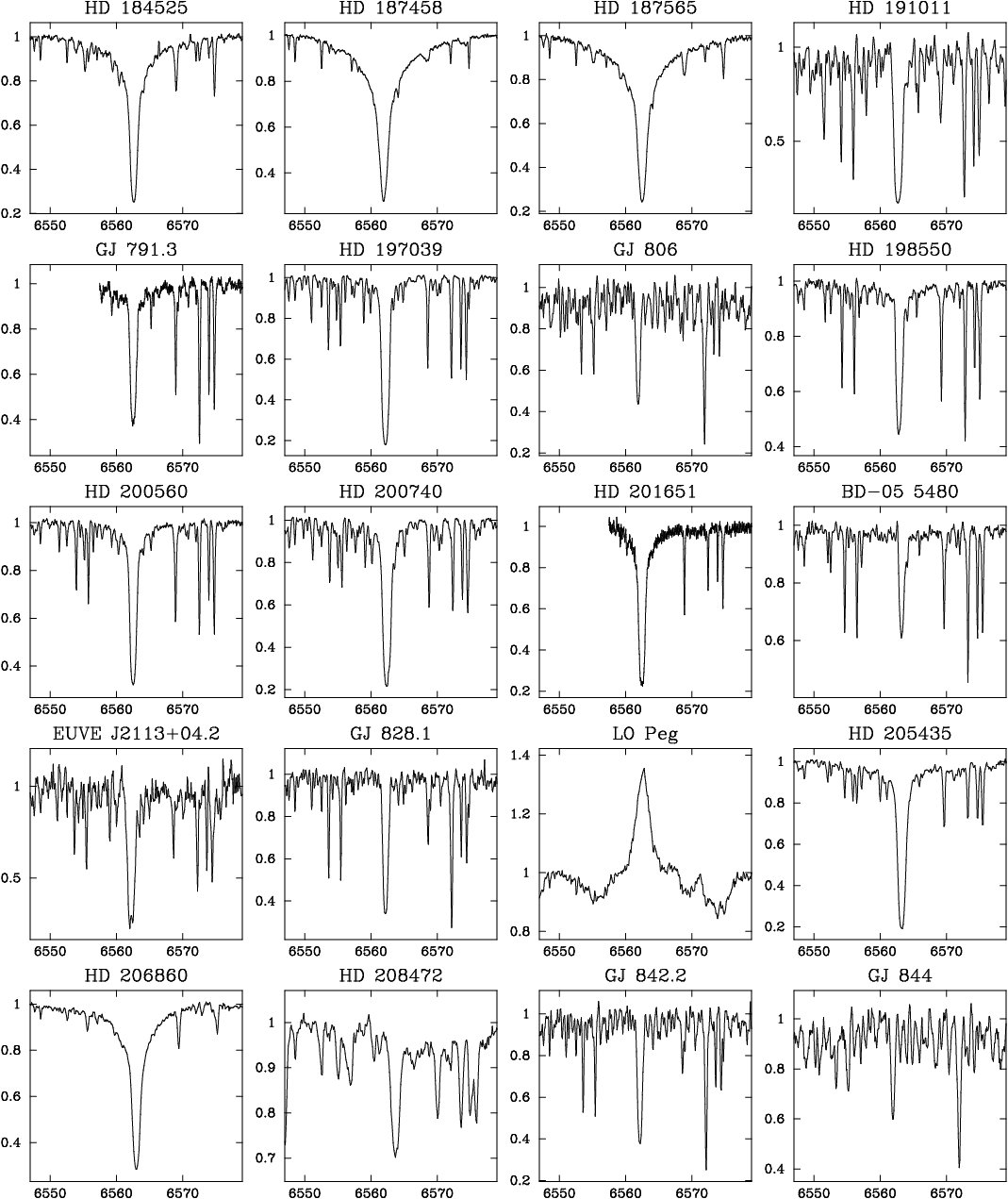}
\caption{Continued.}
\end{figure*}
}
\clearpage
\onlfig{2}{
\begin{figure*}{H}
\includegraphics[]{survey_ha_7.ps}
\caption{Continued.}
\end{figure*}
}

\clearpage
\onlfig{3}{
\begin{figure*}{H}
\includegraphics[]{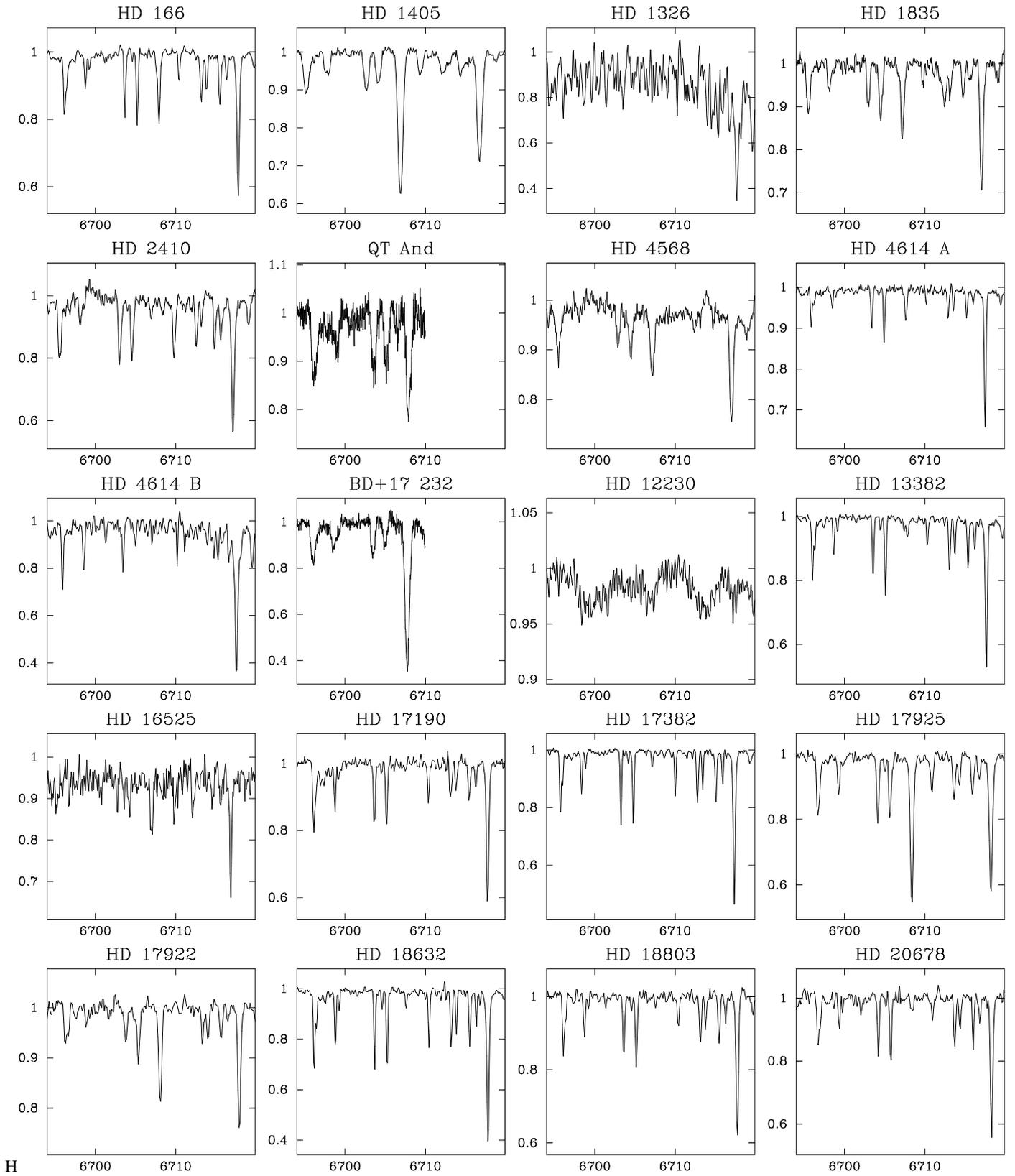}
\caption{\ion{Li}{i} spectra of the stars of our sample with observations 
of this line.}
\label{figA3}
\end{figure*}
}
\clearpage
\onlfig{3}{
\begin{figure*}{H}
\includegraphics[]{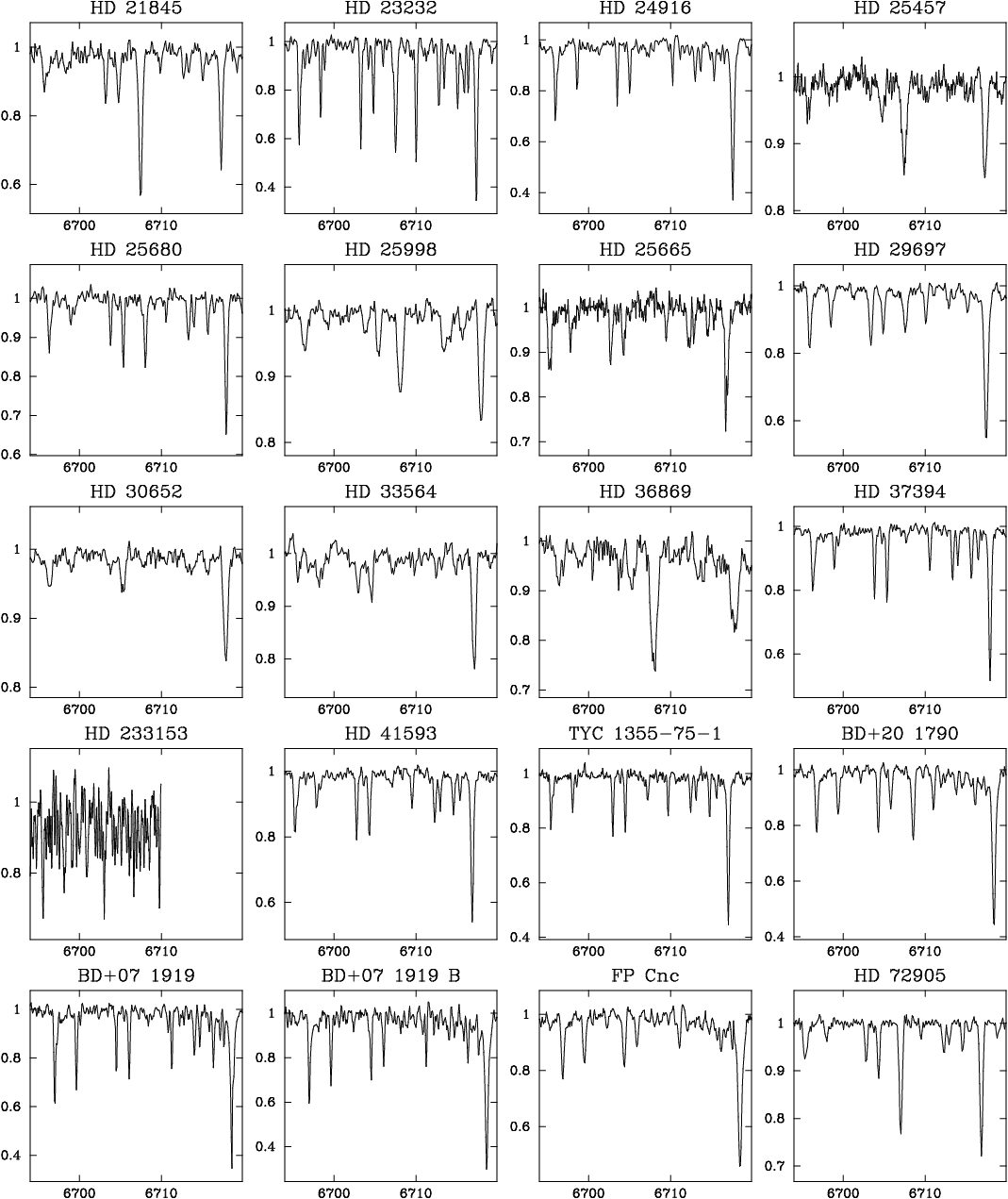}
\caption{Continued.}
\end{figure*}
}
\clearpage
\onlfig{3}{
\begin{figure*}{H}
\includegraphics[]{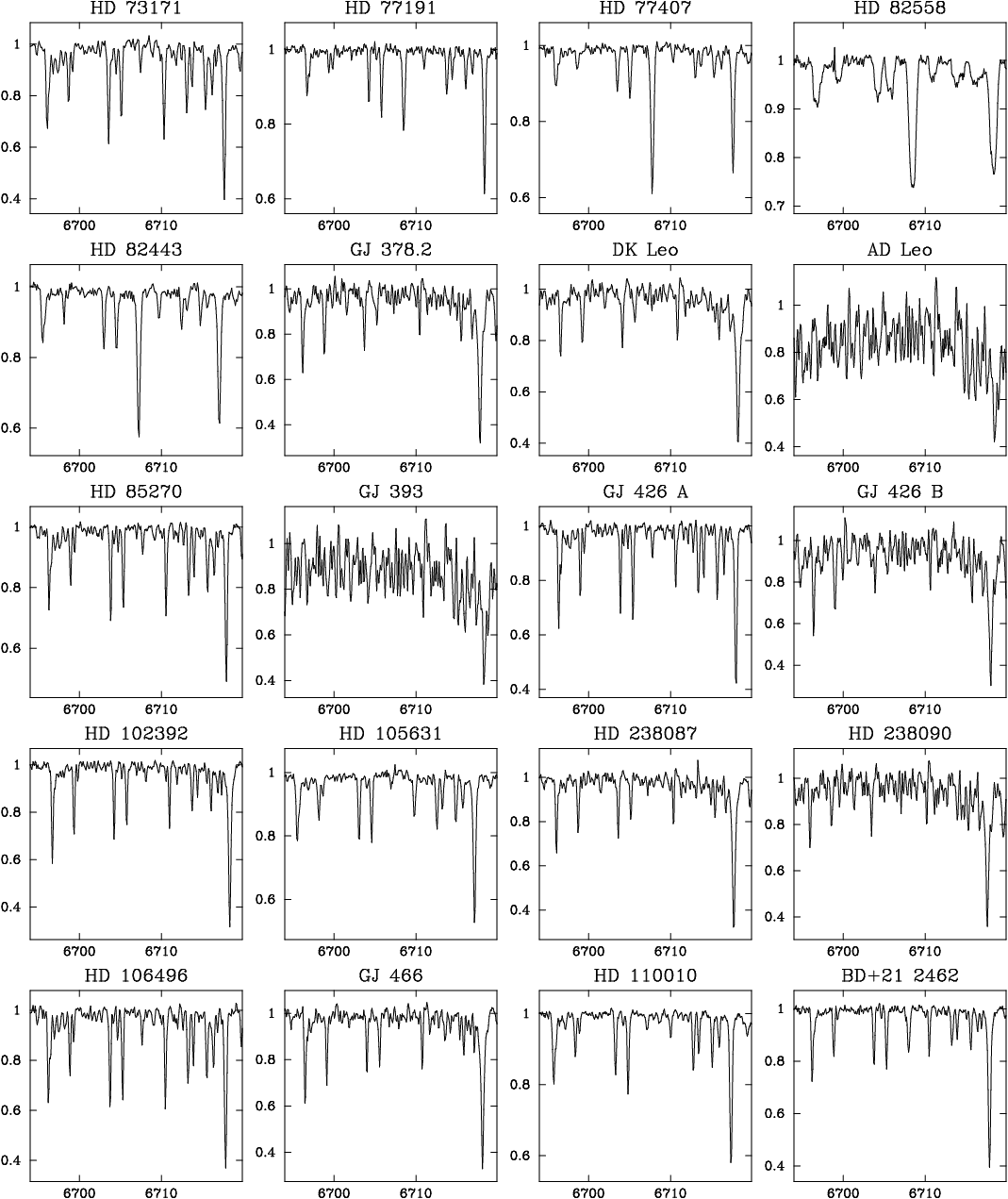}
\caption{Continued.}
\end{figure*}
}
\clearpage
\onlfig{3}{
\begin{figure*}{H}
\includegraphics[]{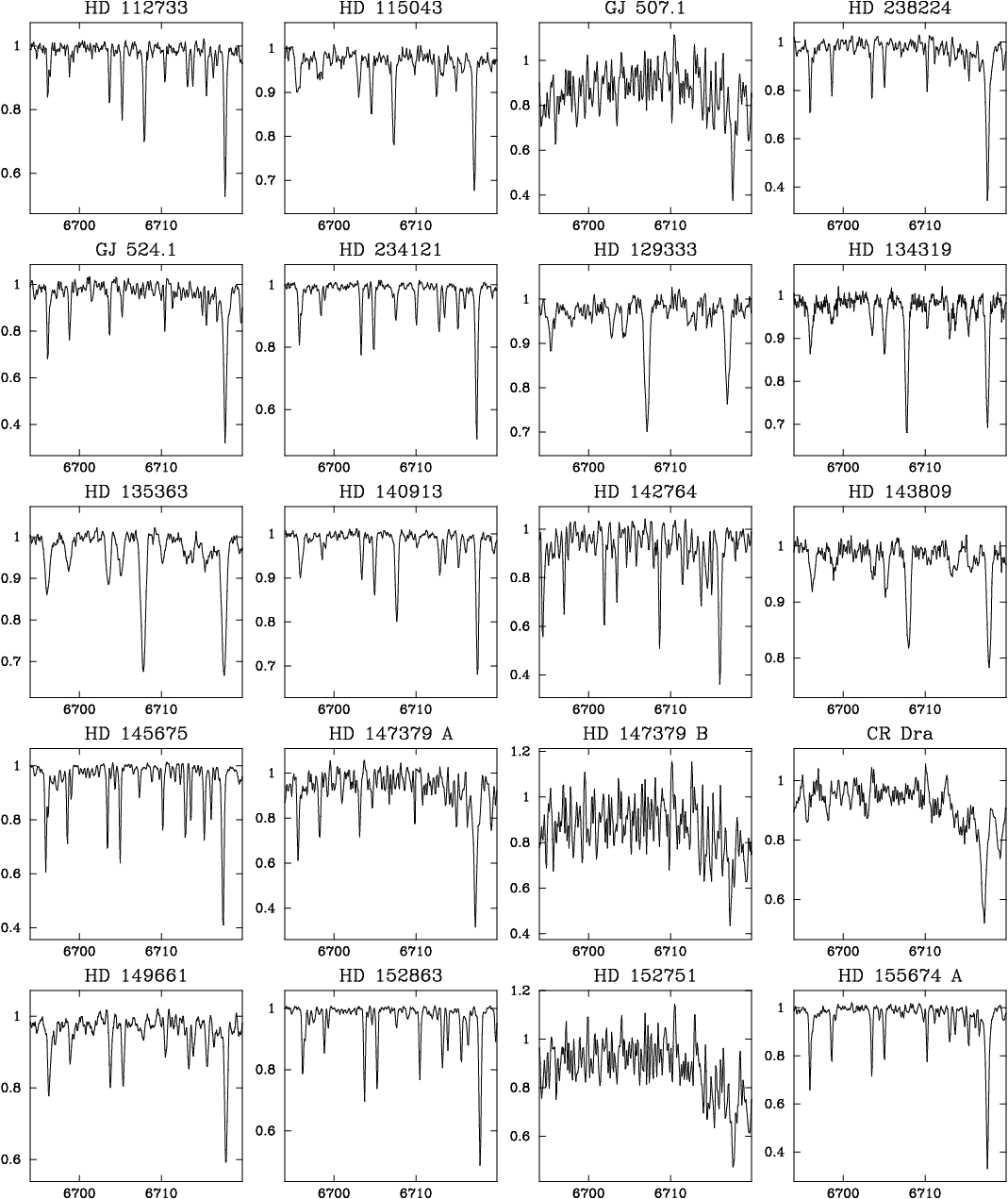}
\caption{Continued.}
\end{figure*}
}
\clearpage
\onlfig{3}{
\begin{figure*}{H}
\includegraphics[]{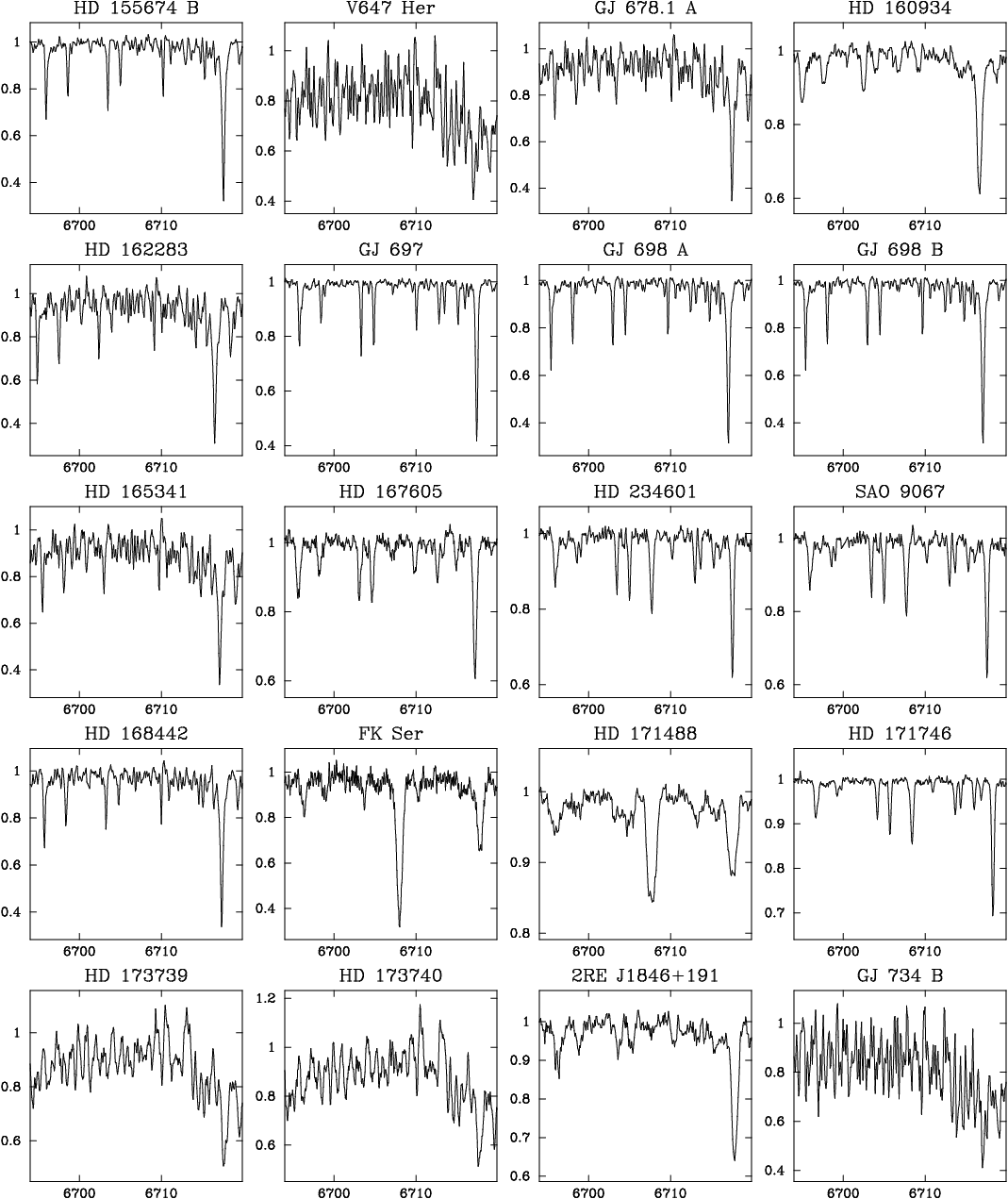}
\caption{Continued.}
\end{figure*}
}
\clearpage
\onlfig{3}{
\begin{figure*}{H}
\includegraphics[]{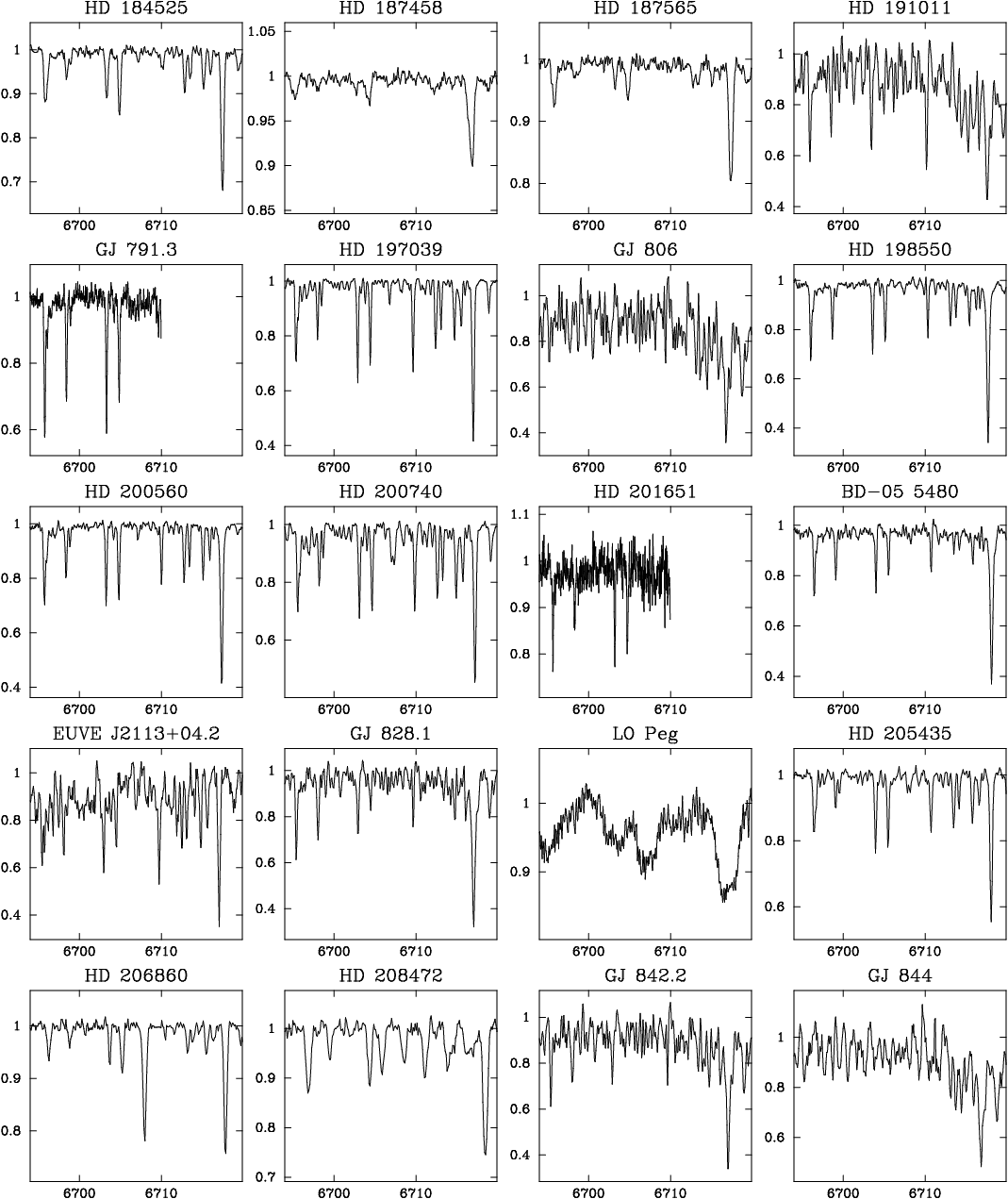}
\caption{Continued.}
\end{figure*}
}
\clearpage
\onlfig{3}{
\begin{figure*}{H}
\includegraphics[]{survey_li_7.ps}
\caption{Continued.}
\end{figure*}
}

\onlfig{4}{
\begin{figure*}
\includegraphics[]{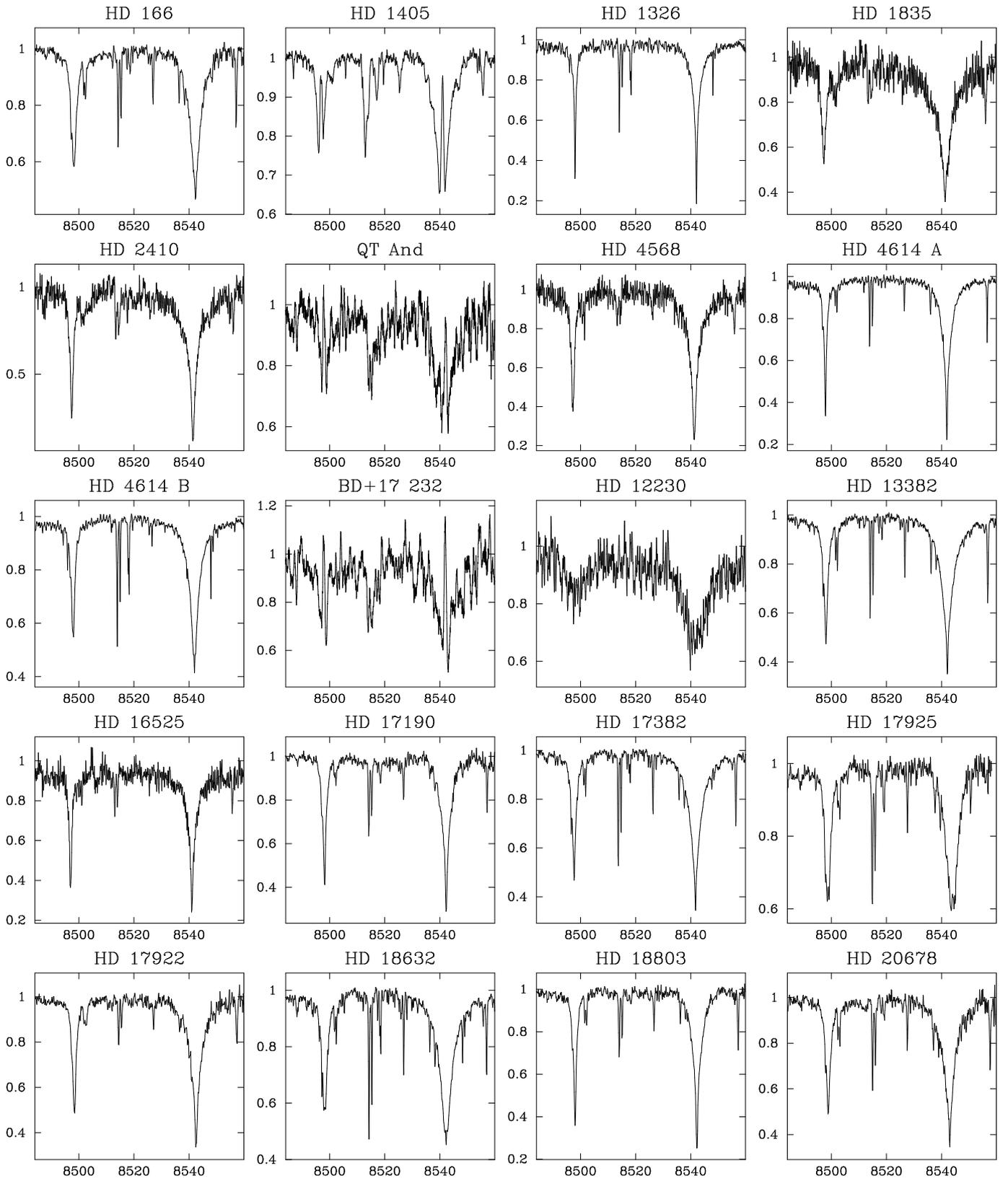}
\caption{\ion{Ca}{ii} $\lambda\lambda$8498 and 8540 \AA\ spectra of the stars of our sample with 
observations of the \ion{Ca}{ii} infrared triplet.}
\label{figA4}
\end{figure*}
}
\onlfig{4}{
\begin{figure*}
\includegraphics[]{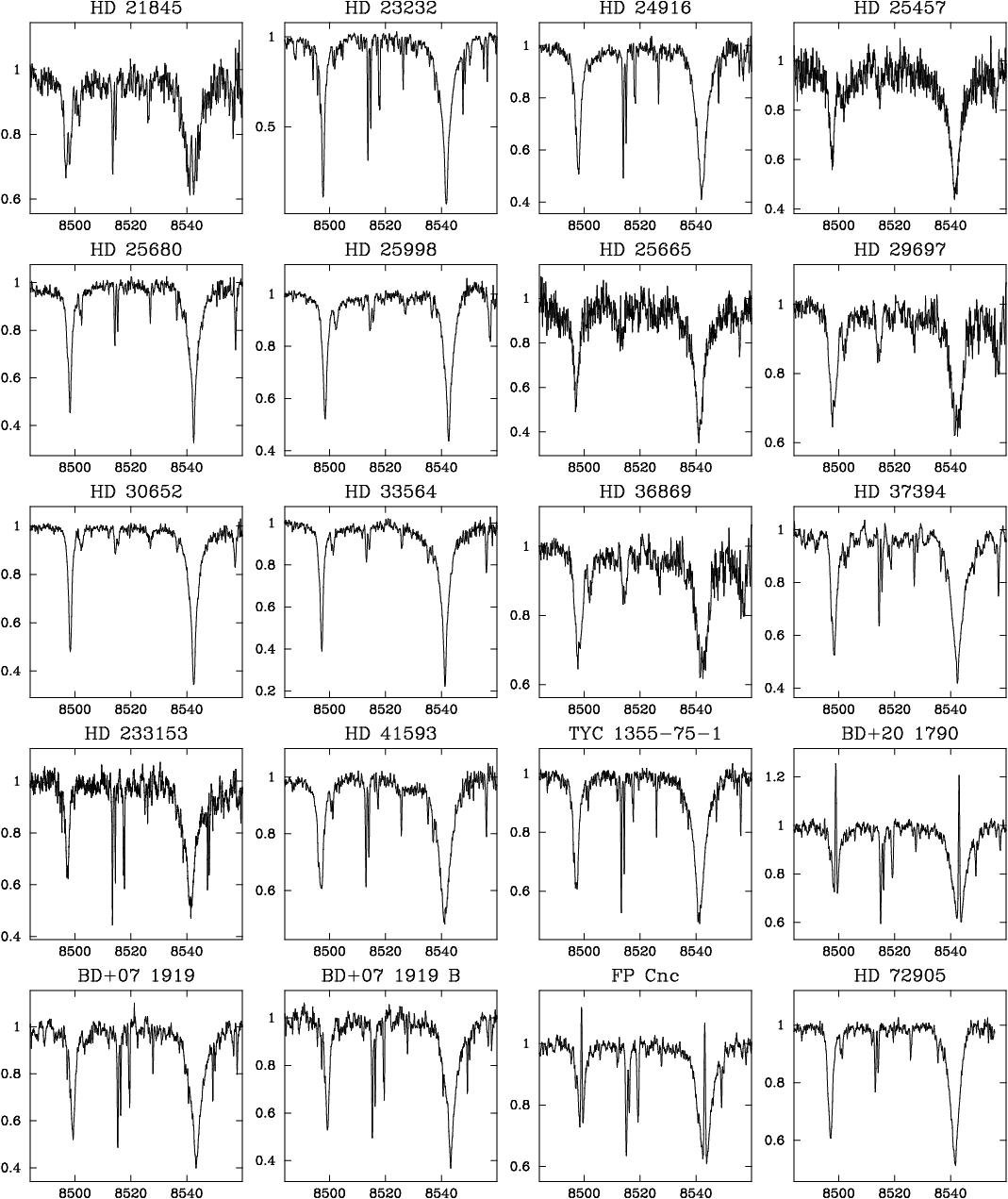}
\caption{Continued.}
\end{figure*}
}
\onlfig{4}{
\begin{figure*}
\includegraphics[]{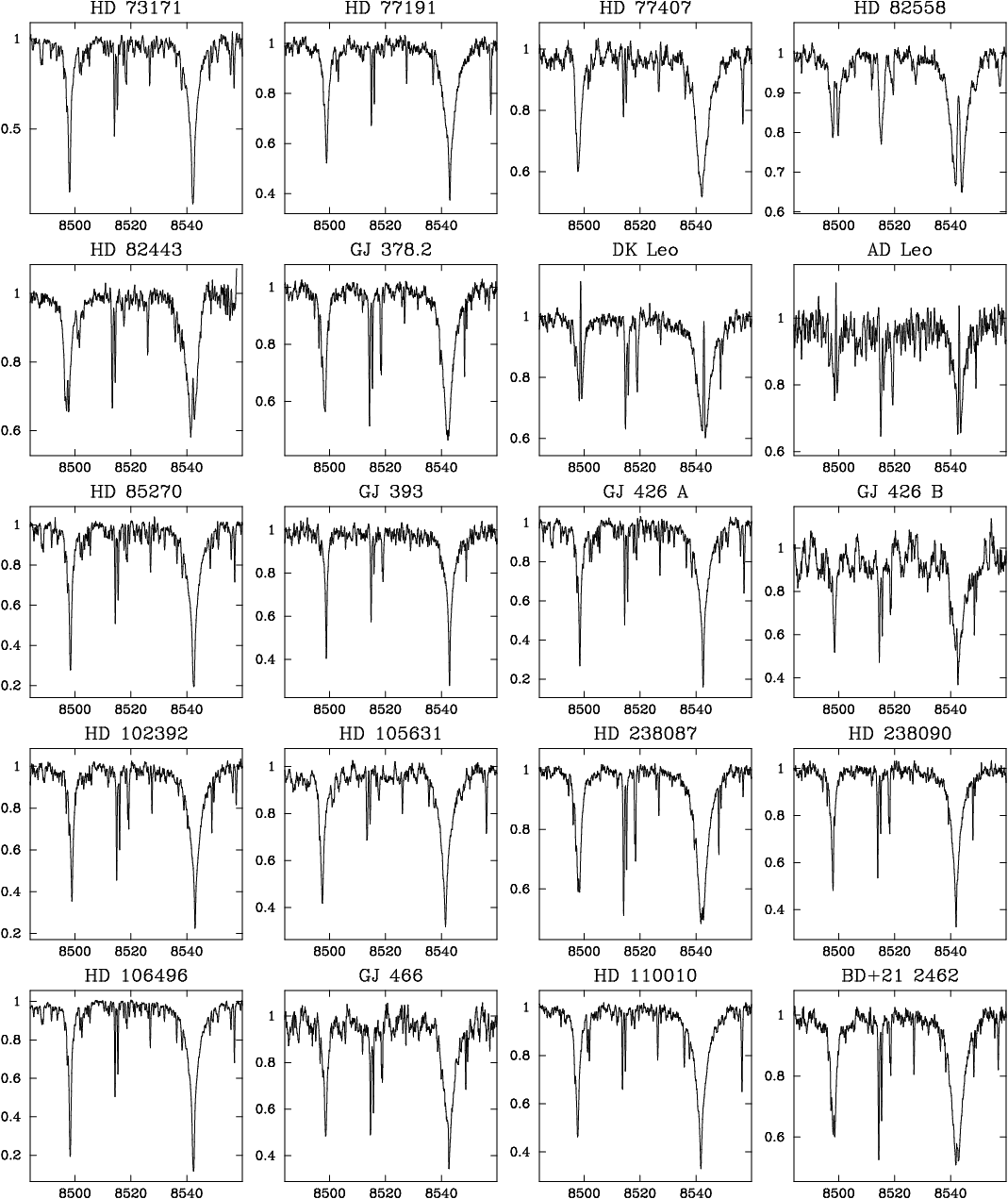}
\caption{Continued.}
\end{figure*}
}
\onlfig{4}{
\begin{figure*}
\includegraphics[]{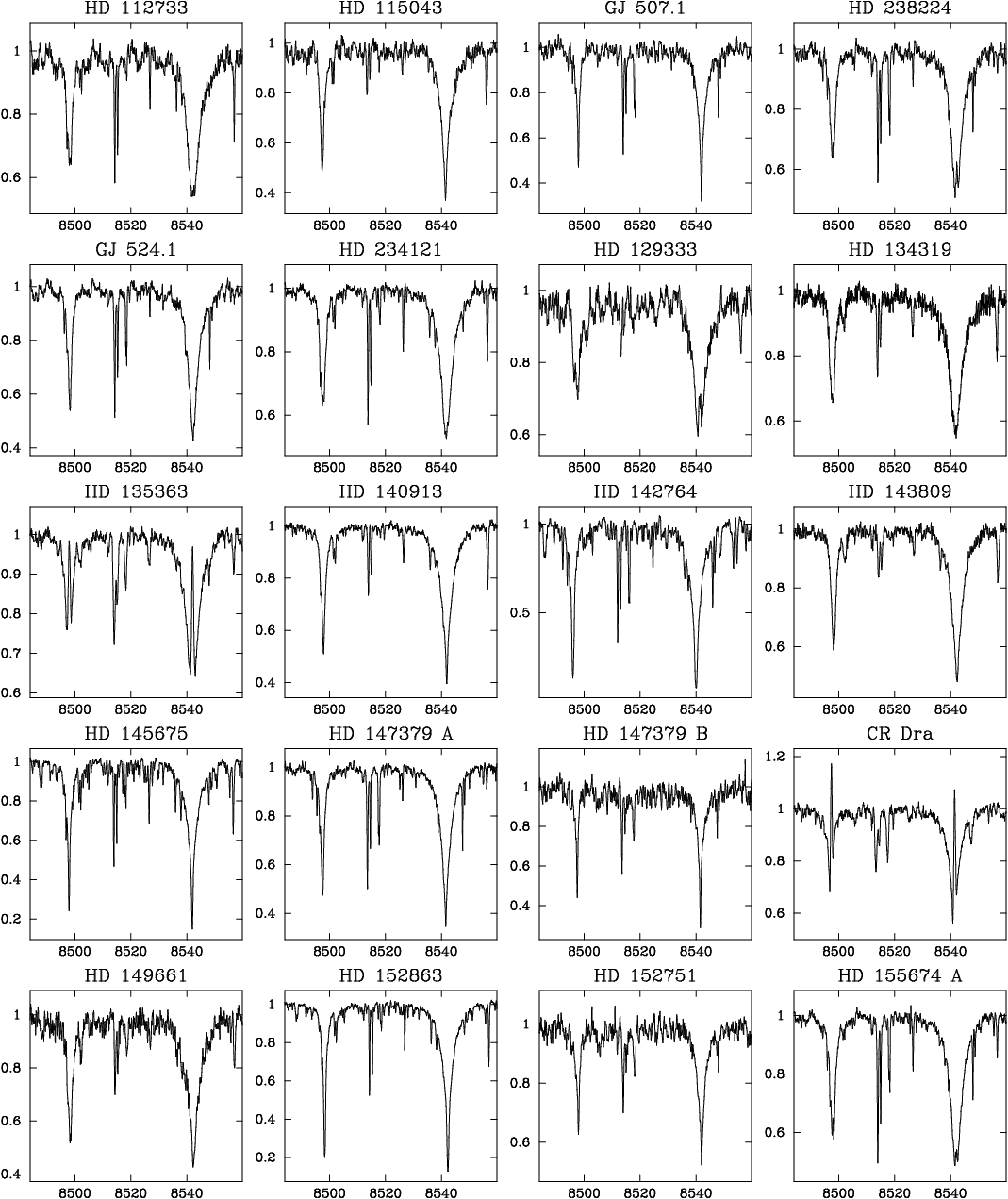}
\caption{Continued.}
\end{figure*}
}
\onlfig{4}{
\begin{figure*}
\includegraphics[]{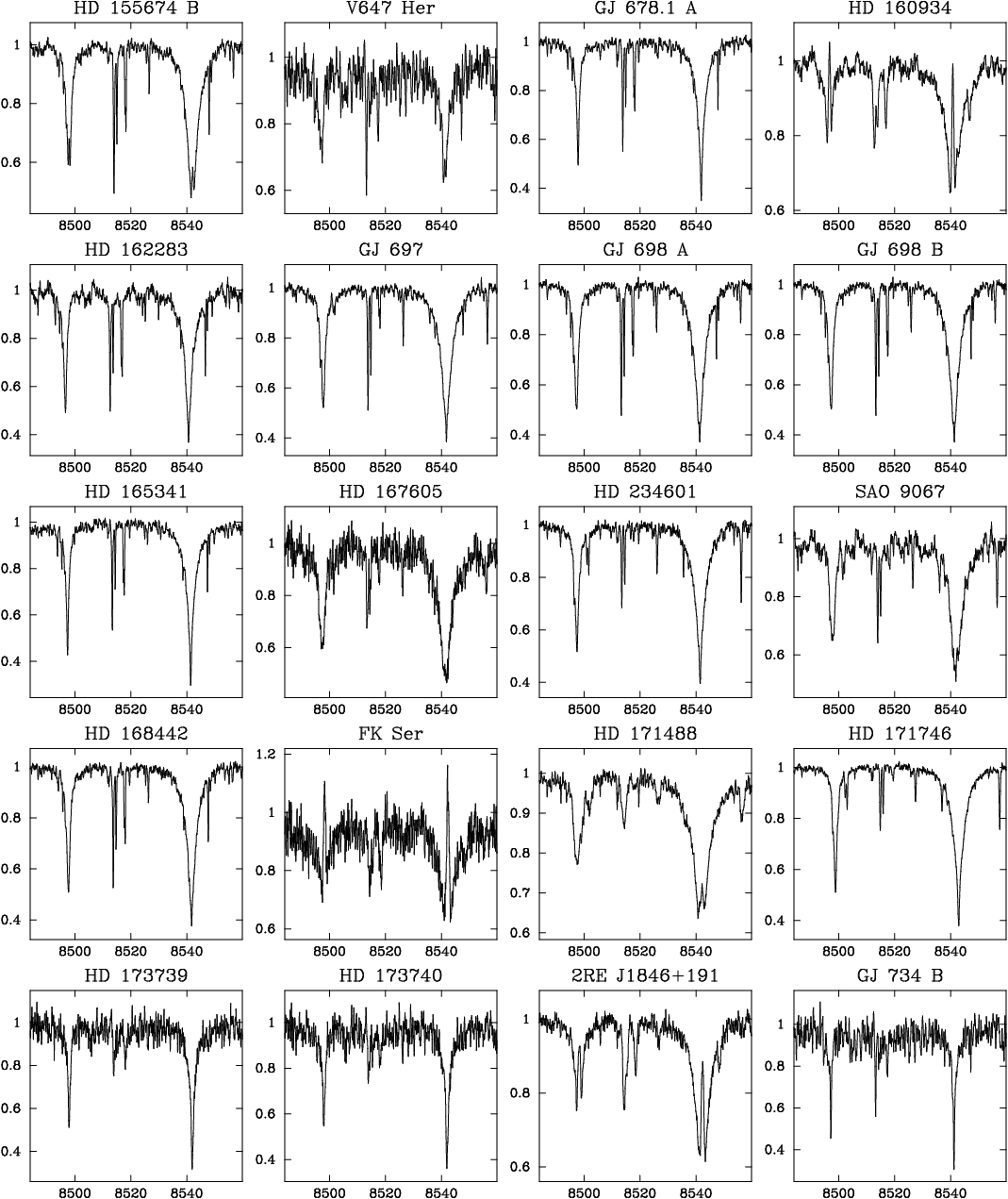}
\caption{Continued.}
\end{figure*}
}
\onlfig{4}{
\begin{figure*}
\includegraphics[]{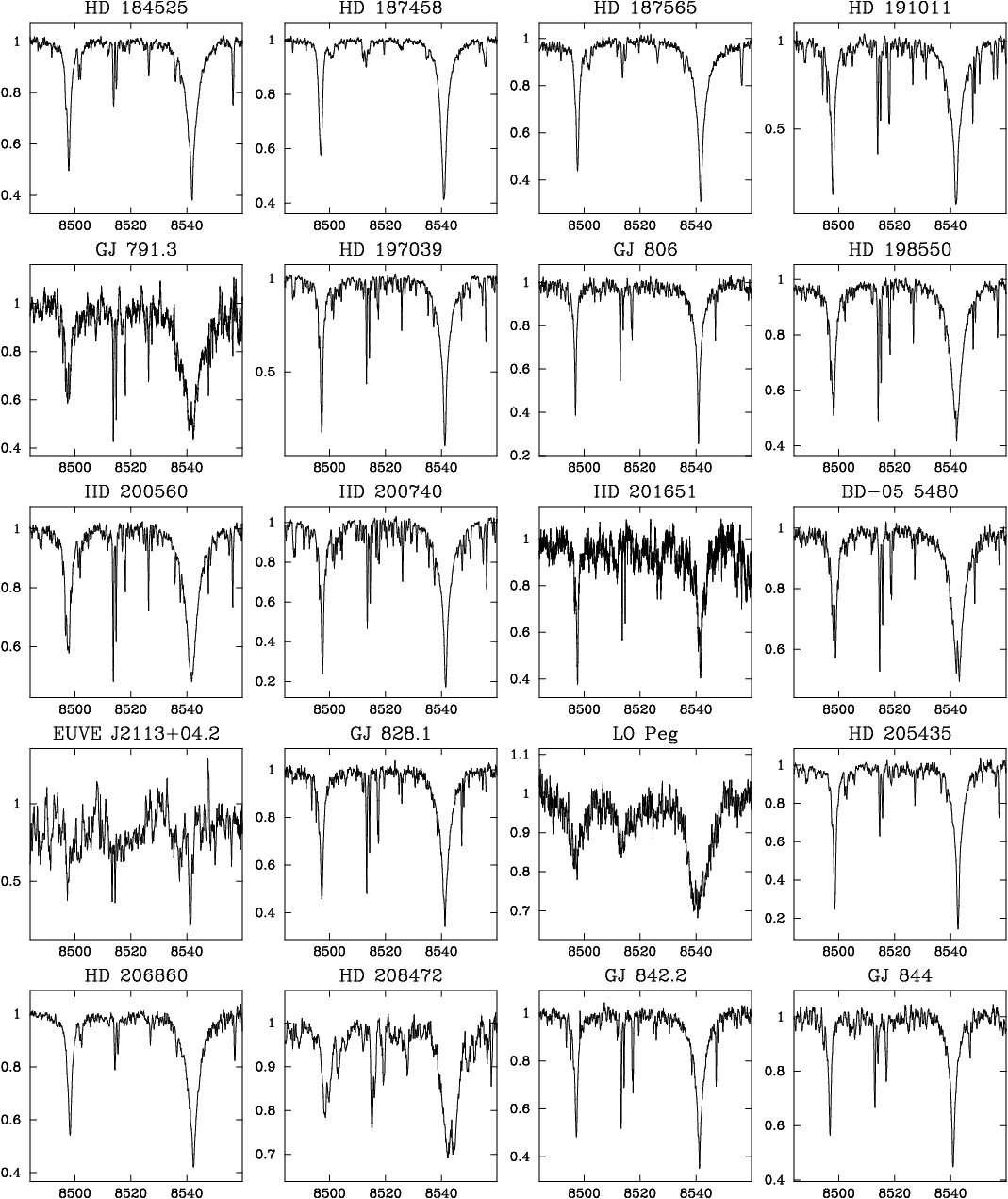}
\caption{Continued.}
\end{figure*}
}
\onlfig{4}{
\begin{figure*}
\includegraphics[]{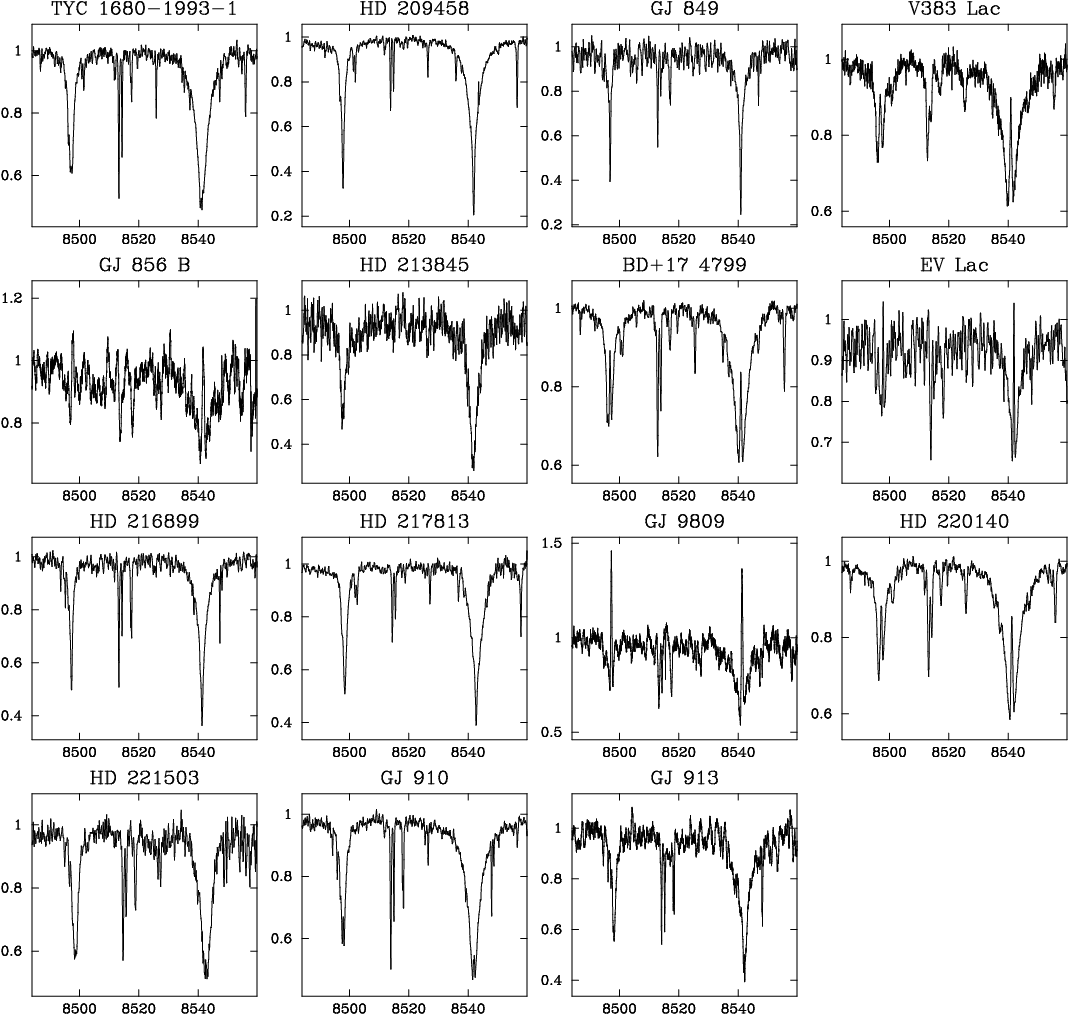}
\caption{Continued.}
\end{figure*}
}

\end{document}